%% file: miniCSC.tex
\documentclass[sn-mathphys,Numbered]{sn-jnl}% Math and Physical Sciences Reference Style
\usepackage{graphicx}%
\usepackage{multirow}%
\usepackage{amsmath,amssymb,amsfonts}%
\usepackage{amsthm}%
\usepackage{mathrsfs}%
\usepackage[title]{appendix}%
\usepackage{xcolor}%
\usepackage{textcomp}%
\usepackage{manyfoot}%
\usepackage{booktabs}%1
\usepackage{algorithm}%
\usepackage{algorithmicx}%
\usepackage{algpseudocode}%
\usepackage{listings}%
\theoremstyle{thmstyleone}%
\theoremstyle{thmstyletwo}%

\theoremstyle{thmstylethree}%

\begin{document}

\title[Article Title]{Longevity Studies of CSC Prototypes Operating with Ar+CO$_{2}$ Gas Mixture and Different Fractions of CF$_{4}$}


\affil[1]{Northeastern University}
\affil[2]{Institute of General and Physical Chemistry, Belgrade}
\affil[3]{University of California Los Angeles}
\affil[4]{University of Florida}
\affil[5]{University of California Davis}
\affil[6]{University of Wisconsin-Madison}
\affil[7]{University of Belgrade}

\author[1]{\fnm{Emanuela  } \sur{Barberis      } }
\author[2]{\fnm{Nebojsa   } \sur{Begovic       } }
\author[1]{\fnm{Nicholas  } \sur{Haubrich      } }
\author[3]{\fnm{Mikhail   } \sur{Ignatenko     } }
\author[4]{\fnm{Andrey    } \sur{Korytov       } }
\author[5]{\fnm{Ota       } \sur{Kukral        } }
\author[4]{\fnm{Ekaterina } \sur{Kuznetsova    } }
\author[6]{\fnm{Armando   } \sur{Lanaro        } }
\author[1]{\fnm{Andrew    } \sur{MacCabe       } }
\author[7]{\fnm{Predrag   } \sur{Milenovic     } }
\author[2]{\fnm{Dubravka  } \sur{Milovanovic   } }
\author[4]{\fnm{Guenakh   } \sur{Mitselmakher  } }
\author[2]{\fnm{Aleksandra} \sur{Radulovic     } }
\author[2]{\fnm{Boris   }   \sur{Rajcic     } }
\author[4]{\fnm{Jake    }   \sur{Rosenzweig } }
\author[1]{\fnm{Bingran }    \sur{Wang      } }
\author[4]{\fnm{Jian    }    \sur{Wang      } }
\author[1]{\fnm{Andrew  }    \sur{Wisecarver} }
\author[1]{\fnm{Darien  }    \sur{Wood      } }
\author[1]{\fnm{Emma    }    \sur{Yeager    } }
%                           
%
%
%
%\author*[1,2]{\fnm{First} \sur{Author}}\email{iauthor@gmail.com}
%\affil*[1]{\orgdiv{Department}, \orgname{Organization}, \orgaddress{\street{Street}, \city{City}, \postcode{100190}, \state{State}, \country{Country}}}
%
%\affil[2]{\orgdiv{Department}, \orgname{Organization}, \orgaddress{\street{Street}, \city{City}, \postcode{10587}, \state{State}, \country{Country}}}
%
%\affil[3]{\orgdiv{Department}, \orgname{Organization}, \orgaddress{\street{Street}, \city{City}, \postcode{610101}, \state{State}, \country{Country}}}


\abstract{
   Studies of Cathode Strip Chamber longevity, comparing Ar+CO$_2$ gas mixtures with fractions of 5\%, 2\%, and 0\% CF$_4$, were performed using several small cathode strip prototype chambers.
   In each trial, a localized source of radiation was used to irradiate up to an accumulated charge of about 300~mC/cm.  Additionally, longevity of a uniformly irradiated prototype operating with 2\% CF$_4$ was studied at the CERN Gamma Irradiation Facility GIF++. Post-hoc analysis of the chamber electrodes using spectroscopy techniques was also done. 
}

\keywords{MWPC, Cathode Strip Chamber, CMS, Longevity, CF$_4$}

\maketitle

\input{Introduction.tex}
\input{PreviousStudies.tex}

\input{Procedure.tex}
\input{LongevityResults.tex}
\input{AnalysisOfElectrodes.tex}

\input{Conclusion.tex}

\bmhead{Acknowledgments}

 We thank the CERN EN-MME-MM group, and especially Ana Teresa Perez Fontenla, for initial set of the material analysis and optical microscopy, CMS mechanical workshop and Valentin Sulimov (PNPI) for help in samples preparation. We are grateful to the CMS CSC group for supporting the studies and to the GIF++ team for help in the irradiation test organization.  We gratefully acknowledge financial support from MoSTDI contract number No.451-03-47/2023-01/200051 (Serbia), DOE (USA) and NSF (USA).
 
\bmhead{Data Availability Statement}
The measurement results and analysis procedures are stored at the common data space of the CMS experiment and can be obtained from the corresponding author on reasonable request.

\bibliography{miniCSC}% common bib file

\end{document}

%% file: Introduction.tex
\section{Introduction and Motivation} \label{introduction}

% Color scheme for writing, editing
% {\color{red}RED: remarks, missing information, or edits to be made} \newline
% {\color{violet}VIOLET: Andrew } \newline
% {\color{orange}ORANGE: Bingran } \newline
% {\color{blue}BLUE: Katerina } \newline
% {\color{black}BLACK: Armando }

%%%%%%%%%%%%%%%%%%%%%%%%%%%%%%%%%%%%%%%%%%%%%%%%%%%%%%%%%%%%%%%%%%%%%%%%%%%%%%%%%%%%%%%%%%%%%%%%%%%%%%%%%%%%%%
% Intro to CSCs as gas-volume chambers of MWPC-like geometry; principle of operation

% Introduction to CMS and CSCs
\subsection{The Cathode Strip Chamber system of the Compact Muon Solenoid}
%The Compact Muon Solenoid (CMS) experiment at CERN's Large Hadron Collider (LHC) is a general-purpose high-energy proton-proton collider experiment designed to perform both precision measurements of standard model phenomena and searches for evidence of beyond the standard model physics. 
%The central feature of the CompactCMS apparatus is a superconducting solenoid of 6\unit{m} internal diameter, providing a magnetic field of 3.8\unit{T}. 
%Within the solenoid volume are a silicon pixel and strip tracker, a lead tungstate crystal electromagnetic calorimeter (ECAL), and a brass and scintillator hadron calorimeter (HCAL), each composed of a barrel and two endcap sections. 
%Forward calorimeters extend the pseudorapidity coverage provided by the barrel and endcap detectors. 
The CMS detector features a powerful system for muon detection. The base component of this system in forward ``endcap'' regions (0.9$<|\eta|<$2.4) comprises 540 Cathode Strip Chambers (CSCs), which are  multiwire proportional chambers~\cite{Sauli:1977mt}. The CSCs are used for muon identification, triggering, and track measurements~\cite{CMS:2008xjf}. The chambers are arranged in 4 muon stations located between steel disks of each CMS endcap. CSCs have trapezoidal shape and are grouped in rings. A muon station consists of 2 or 3 rings located at different radial distances from the LHC beam pipe. Rings of CSCs are labeled as ME$S/R$, where $S$ indicates the station position in $|z|$ and $R$ gives the ring number. All chamber types, except for the most forward ME1/1~\cite{Ershov:2006sf}, differ only in the size of the active area and are constructed of the same material using the same assembly technology. The CSC system was installed before the startup of the LHC, except for ME4/2 chambers, which were produced and installed during the LHC long shutdown in 2013-2014. The CSCs have operated successfully since the beginning of CMS operations, and continue to do so. A future upgrade of the accelerator complex to High-Luminosity LHC (HL-LHC) will increase luminosity to reach (5-7.5)$\times10^{34}$ cm$^{-2}$s$^{-1}$. After the readout electronics upgrade performed during the LHC long shutdown 2018-2021, the CSC system is intended to continue efficient operation through the HL-LHC era.

% {\color{orange} Also shown in Fig.~\ref{fig:cscsignalcartoon} are smaller size CSCs (Fig.~\ref{fig:cscsignalcartoon}, right), hereafter referred to as miniCSCs, which have been constructed and used for this investigation. The design of miniCSCs will be discussed in Section \ref{procedure904miniCSCDesign}.}

\begin{figure}[htbp]
\begin{center}
\includegraphics[height=0.4\textwidth]{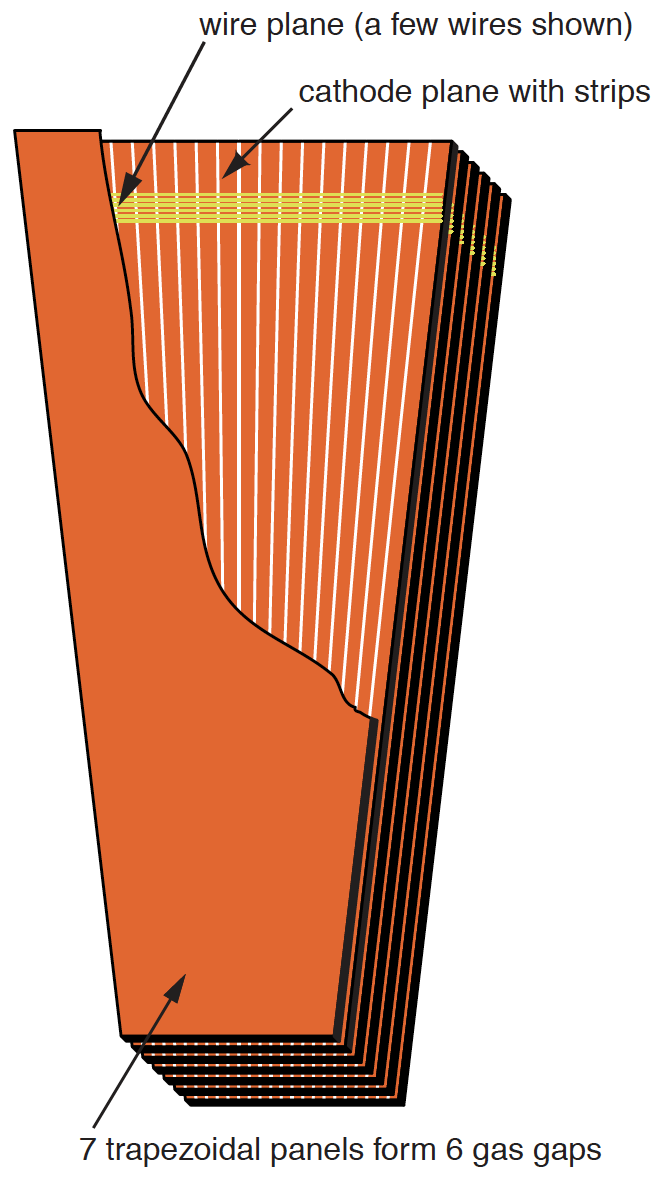}
\includegraphics[height=0.4\textwidth]{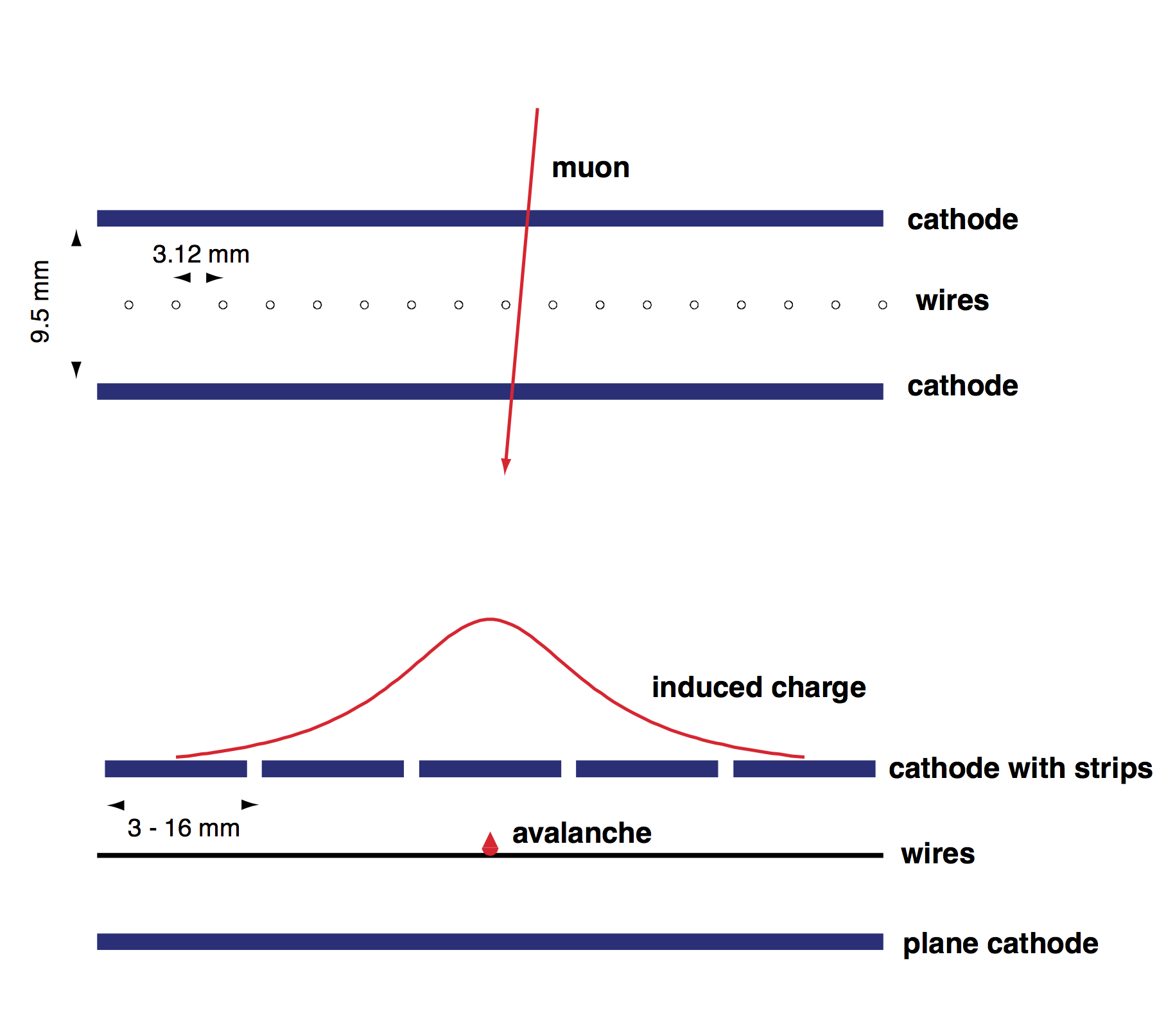}
\caption{The construction of a full-sized CSC, with part
of the top layer cut away to show part of the wire plane (in gold) with the cathode strips below (left) and the depiction of signal induction on the cathode strips created by an ionizing particle traversing the chamber volume (right).
    \label{fig:cscsignalcartoon}}
\end{center}
\end{figure}

Each CSC consists of six detecting layers with two cathode planes made of copper-cladded FR4~\cite{Layter:343814}. One of the cathode planes has the copper coating milled into strips of several millimeters width (Fig.~\ref{fig:cscsignalcartoon}, left) so the charge induced on the cathode is shared among several readout channels (Fig.~\ref{fig:cscsignalcartoon}, right) in order to provide a good track spatial resolution in azimuthal plane. Radial track coordinates are measured using anode signals.  Gold-plated tungsten anode wires have 50~$\mu$m diameter and are perpendicular to the cathode strips for the non-ME1/1 CSC types. 
%More detailed description of the CSC parameters is given in Table \ref{tab:miniCSCspecs} of Section \ref{miniCSCDesign}. 

The CSC gas mixture contains 40\% Ar, 50\% CO$_{2}$, and 10\% CF$_{4}$ and provides stable chamber operation in the proportional mode with the average gas gain around 6x10$^4$.

%Each of these components serves a different purpose. 
%The argon gas is chemically inert and  non-reactive, with an ionization potential at 15.8 eV~\cite{lide2004crc}, and it is the primary component that is ionized by the incoming particles. 
The relative fraction of Ar in the mixture controls the charge amplification of the initial energy deposit (gas gain) and consequently the minimum voltage required to operate the chamber with a reasonable gas gain.  
%This voltage corresponds to the ``knee'' of the ``efficiency plateau'', in which the chamber operates in proportional mode and is able to efficiently count physical ionizing particles that traverse the gas volume.
CO$_{2}$ is primarily used for its quenching effect that allows the detection volume to recover after the passage of an ionizing particle by absorbing high-energy photons produced in the avalanche. 
This shortens the ``dead-time'' after the initial signal has passed and protects the chamber from damage due to sparking, caused by a positive feedback loop of secondary and tertiary reactions. 

Adding 10\% CF$_{4}$ to the Ar/CO$_2$ mixture extends the chamber lifetime preventing anode wire aging~\cite{FERGUSON2002240}. 
%In terms of operation, the fraction of CF$_{4}$ is related to the voltage range where the chamber can efficiently detect incoming signals (the length of the efficiency plateau). 
%For higher voltages within this efficiency plateau, CF$_{4}$ acts as a quencher to suppress spurious signals. 
%\FIXME{in the CSC geometry CF4 changes drift time only starting from ~40\%. below is the updated text. }
Though CF$_{4}$ also increases the drift velocity of the electrons in a gas mixture, this effects depends on the electrical field configuration and for the CSC geometry is negligible if the CF$_{4}$ fraction in the gas mixture is below 20\%~\cite{Kisselev:1997cba}. Thus at the nominal CSC operation voltages the main purpose of CF$_{4}$ is extension of the chamber longevity.

%with a maximum drift time of xx ns for 10\% CF$_{4}$, compared to xx ns for 0\% CF$_{4}$.
% CF$_{4}$'s role in internal plasma chemistry, prevention of aging}

\subsection{Chamber aging}

In a CSC the avalanche charge amplification of several ten thousand times occurs in vicinity of a thin anode wire. The avalanche can be considered as short plasma discharge of microscopic size where ions and free radicals of the gas atoms and molecules are produced, with some radicals having sufficient lifetime to reach CSC electrodes. Plasma chemical reactions may cause modification of the chamber electrode surfaces either due to direct reactions of the plasma products with the electrode material, or due to formation of deposits on the electrodes. This may lead to degradation of the chamber muon detection performance and so may limit its long-term operation. 

The most dangerous type of aging is modification of the anode surface. Any deposit on the anode wire may increase its diameter causing non-uniform reduction of the gas gain and, in case of the cathode strip chambers, a degradation of the spatial resolution. Upon significant decrease of the gas gain, the induced signal may become too low to provide efficient muon detection with the nominal readout electronics.  Sharp needle-like deposits cause increase of the single layer noise rate and, in the worst cases, may result in appearance of sparks or even electrical breakdowns. 

Thin insulating deposits on cathode surface may cause increase of secondary electron emission resulting in high noise rate or even in rather strong persistent currents (so-called Malter effect~\cite{PhysRev.50.48, Albicocco:2019jvz}). If cathode deposits are conductive and cover gaps between CSC strips, they may affect the spatial resolution. 

At the same time, the plasma conditions may be used to protect the chamber against degradation, as it is done using a small quantity of CF$_4$ in the working gas mixture of the CMS CSCs. Fluorine radicals created in the plasma discharges react with silicon or carbon -based molecules, which may be present in various quantities in chamber material or gas impurities, and are proven to provoke aging~\cite{Kadyk:1990uq, Vavra:2002tds}.
Fluorine radicals prevent polymerization of organic molecules and also bond silicon into volatile SiF$_4$ molecules, which are removed from the chamber with the gas flow. 

% {\color{red}(Note: insert cartoon from Vavra aging paper to show illustration of chemistry?)}

% Motivation to reduce CF$_{4}$ use
\subsection{Motivation for reducing or eliminating CF$_{4}$}
%The primary motivation for reducing CF$_{4}$ use is environmental, secondary is financial;
CF$_{4}$ is an expensive gas and its release into the atmosphere is detrimental to the environment.  
It belongs to a class of fluorocarbon gases that have a negative impact on atmospheric chemistry, leading to greenhouse warming effects on the Earth's surface.
To be specific, CF$_{4}$ has a Global Warming Potential (GWP) of 6630 over 100 years~\cite{IPCC_CH8}. 

%\FIXME{Due to its high GWP, it is likely that the 
%European Union may impose restrictions on its consumption or release into the atmosphere.}
%\FIXME{new text:}\\
CF$_{4}$ is subject to the European Union F-gas regulation (Regulation (EU) No 517/2014,~\cite{f-gas_regulation}) which aims to the reduction of the F-gas emission by two-thirds by 2030 compared with 2014 levels.
Moreover, recently the European market started to experience periodic issues in supply chains that makes CF$_{4}$ more difficult and expensive to obtain  regularly. 
So, either lowering the fraction of CF$_{4}$ or excluding it entirely from the CSC gas mixture is well-motivated. 

One of the approach for reducing the CF$_{4}$ exhaust is recirculating and recuperating the gas. Recirculation was implemented from the beginning of the CSC operation using the closed-loop gas system with 10\% replenishment rate. Additionally, a CF$_{4}$ recuperation system was developed during last decade by the CERN EP-DT group and brought to operation with about 60\% efficiency by the beginning of Run3.

In addition to that, studies of the possible reduction of the CF$_{4}$ content in the CSC gas mixture were started.

%% file: PreviousStudies.tex
\section{Previous Studies of CSC gases}

% {\color{red}(Point of this section is to re-state the history of why CF$_{4}$ was thought necessary for chamber longevity; original longevity tests with CSCs; why we are verifying CF$_{4}$'s use in longevity again)}

In the early stages of design and testing of the CSCs, before their initial installation in CMS, the longevity of chambers was  studied~\cite{FERGUSON2002240} by operating test chambers while being radiated by a strong $^{90}$Sr source. 
These chambers were constructed from mostly the same materials as those intended for the production chambers.  
A notable exception was the use of silicon-based RTV sealant first applied before the chamber was closed, allowing a possible leak of RTV inside the chamber, creating a source of silicon in the gas volume.

For two gas mixtures containing CF$_{4}$, Ar/CO$_{2}$/CF$_{4}$ 40\%/50\%/10\% and 30\%/50\%/20\%, the irradiation was continued up to a local accumulation of about 13 C/cm on the anode wire. 
Even with this large accumulated charge, no appreciable decrease was observed in the gas gain. Some increase of the dark current was observed, which indicated aging effects on the cathodes, but the overall chamber performance was not degraded significantly.

For a 70\%/30\% Ar/CO2 mixture without CF$_{4}$, however, a gas gain drop of a factor of two was observed for each 0.25 C/cm accumulated.  
In a post-mortem examination of the test chambers, a thick coating was found on the anode wires.  
An elemental analysis of this coating revealed a major component of silicon.  The source of the silicon was likely the RTV, and the probable conclusion was that the CF$_{4}$ was effective in preventing the build-up of silicon-based polymers on the anode wires.

%The conclusion of these studies was that a fraction of 10\% CF$_{4}$ in the gas mixture was sufficient in order to prevent severe aging in CSCs, and this is the fraction that was selected for operation in Run 1 and Run 2.  
%Several caveats accompanied this study. The irradiation was highly local and so assumptions were required to extrapolation to full-chamber irradiation. The acceleration factor was very large; the charge expected in many years of LHC running was accumulated in a few weeks.  Finally, RTV was applied in the test chambers before closing while in  the production chambers it was applied afterwards to prevent leakage into the gas volume.

In the years 2000 and 2001, two full-size chambers were studied
for aging~\cite{Prokofev:2001zp} using the 740 GBq
radioactive $^{137}$Cs source (0.662 MeV gammas) at the Gamma Irradiation Facility (GIF) at CERN.  The gas mixture used was Ar/CO$_{2}$/CF$_{4}$ 40\%/50\%/10\%, with an open-loop gas system in 2000 and a closed loop system in 2001. The accumulated charges were 0.3-0.4 C/cm. In both cases, no appreciable decrease in gain was observed.
%
%Additional studies~\cite{Anderson:837542} were carried out to explore three different mixtures of Ar/CO$_{2}$/CF$_{4}$:  40\%/50\%/10\%, 60\%/\%30/\%10, 40\%/3\%/5\%.
%Properties such as gain and voltage plateau were investigated, but aging was not studied.
%The authors suggested that 5\% CF$_{4}$ might also be suitable and have cost advantages, but warned that the longevity properties needed to be studied.

In anticipation of the HL-LHC, 
a new campaign of aging studies was launched~\cite{WANG2020162279} with full-sized spare CSCs operated with the nominal gas mixture Ar/CO$_{2}$/CF$_{4}$ 40\%/50\%/10\%. The irradiation was performed at the upgraded Gamma Irradiation Facility (GIF++) with 14 TBq $^{137}$Cs source~\cite{Jakel:2014hxx}. 

The gain and resolution of the chambers were monitored and showed no sign of deterioration up to the accumulated charge of 0.33~C/cm, which corresponds to the charge expected at the HL-LHC conditions with a safety factor of at least 1.5, depending on the chamber type and its position in the CMS detector.  
%Note that subsequent updates to the designs of end cap calorimeter and the beam pipe resulted in increased estimates for the background rates in the end cap region, 
%so the safety factor for these studies is now known to be less than three.

% List of sections, final introductory statement
%\subsubsection{Aging tests for operation at the HL-LHC}
%The expected increase in integrated luminosity and higher integrated charge require higher qualification standards for the detectors. Thus, CSCs were submitted for a second round of aging tests to verify the detectors would still operate after three times the integrated charge expected after ten years of operation at the HL-LHC expected instantaneous luminosity of 5~$\times$~10$^{34}$~cm$^{-2}$s$^{-1}$.

%% file: Procedure.tex
\section{Experimental Design and Procedure} \label{procedure904}

As detailed in Section \ref{introduction}, aging can be defined as a significant chemical change on the surface of electrodes inside the detection volume of a gaseous proportional chamber. 
Depending on the type of chemical reactions occurring, operation stability, gas gain, efficiency or resolution may be affected; these parameters can be monitored through non-destructive techniques. 
Additionally, if the chambers can be disassembled after exposure, spectroscopic techniques can be used to study the change of the electrode material itself.

To allow rapid and flexible tests, and to facilitate post-mortem analysis of the wires and cathode surfaces, a set of small CSC prototypes, or ``miniCSCs'', was constructed.
The aging tests with the miniCSCs were carried out in three locations: the CERN 904 lab (B904), the GIF++ radiation facility at CERN, and the Petersburg Nuclear Physics Institute (PNPI) in Russia.

% It is thus necessary to chose a set of basic, low-level observables as a handle to monitor the ``health'' of the chamber as a function of the aging. 

% In Section \ref{procedure904miniCSCDesign} and \ref{procedure904testsetup}, details of the miniCSC construction and the setup of the test stand used to take measurements are described. The observables monitored over the course of the study, the motivation for choosing them, and how they are measured, is described in Section \ref{procedure904studiedobservables}.

%%%%%%%%%%%%%%%%%%%%%%%%%%%%%%%%%%%%%%%%%%%%%%%%%%%%%%%%%%%%%%%%%%%%%%%%%%%%%%%%%%%%%%%%%%%%%%%%%%%%%%%%%%%%%%
% Design specifications for the miniCSC chambers

%%%%%%%%%%%%%%%%%%%%%%%%%%%%%%%%%%%%%%%%%%%%%%%%%%%%%%%%%%%%%%%%%%%%%%%%%%%%%%%%%%%%%%%%%%%%%%%%%%%%%%%%%%%%%%
% Design specifications for the miniCSC chambers

\subsection{miniCSC chamber design} \label{miniCSCDesign}
 
\begin{figure}[htbp]
\begin{center}
\includegraphics[height=0.4\textwidth]{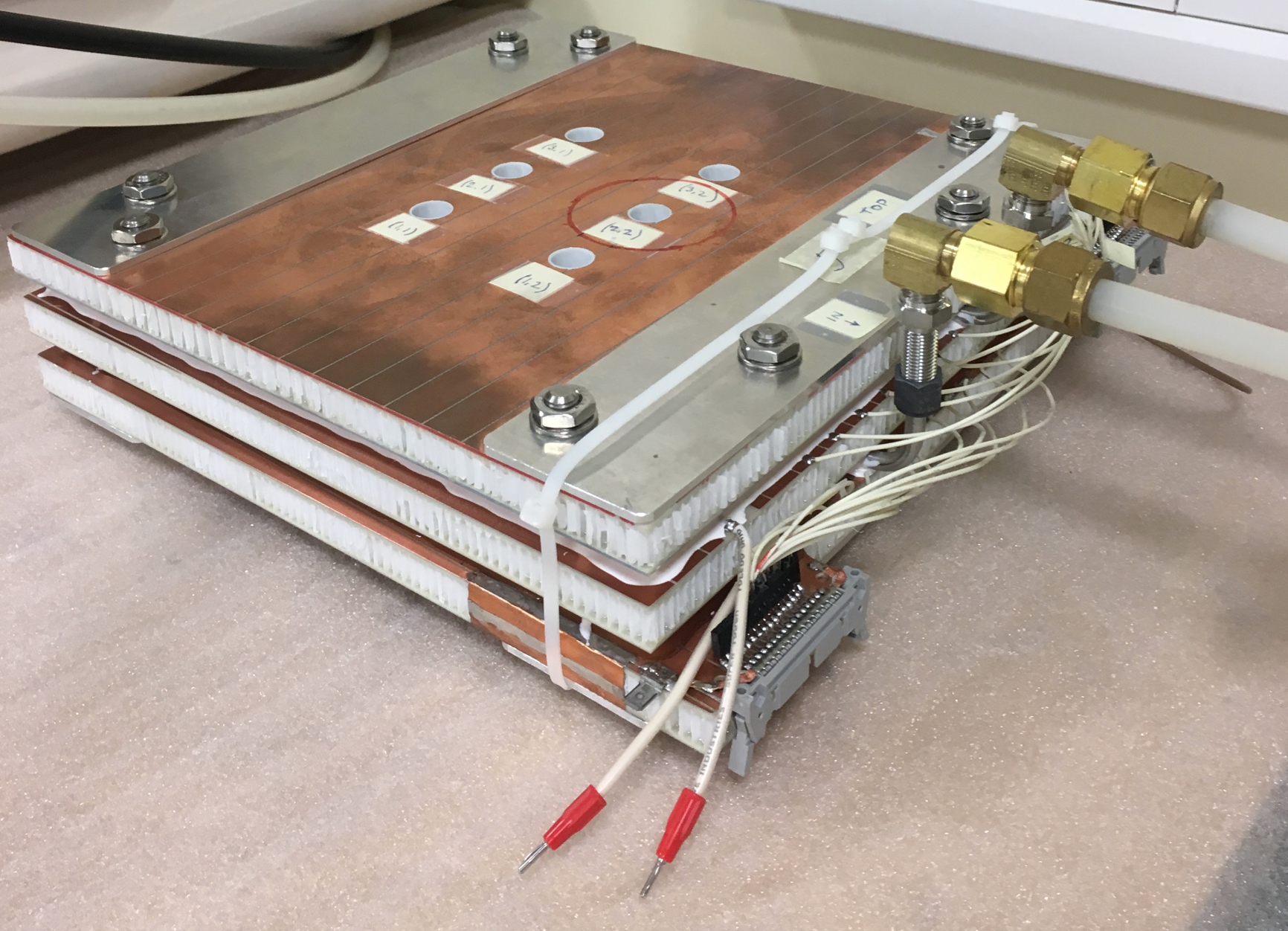}
\caption{Front view of a miniCSC chamber.
    \label{fig:miniCSCpictures_noSource}}
\end{center}
\end{figure}

%A miniCSC is a smaller version of one of the full-size CSCs used in the CMS detector, and an example is shown in Fig.~\ref{fig:miniCSCpictures_noSource}.
A miniCSC is a small prototype of the full-size CMS CSC, and an example is shown in Fig.~\ref{fig:miniCSCpictures_noSource}.
These chambers were built by using the same materials as used for the construction of the ME4/2 CSCs in 2012-2013, following the same assembly procedures and quality checks. 
The dimensions of the miniCSCs are 30~cm$\times$30~cm, and they have two layers instead of six, with a volume of about 0.5 L for a single gas gap.
As for the standard CSCs, the cathodes of the miniCSCs are composed of copper-clad FR4 with strings milled into the copper, and the wires are gold-plated tungsten of 50~$\mu$m diameter.  
The geometry of the chamber was inherited from the original CSC and is specified in Table \ref{tab:miniCSCspecs}.
To provide efficient irradiation and allowing control measurements with a $^{109}$Cd X-ray source, cylindrical holes were drilled into the FR4 of the outer cathode layer, still preserving the integrity of the cathode surface. 

%In most cases, the components came from material left over from the construction of the ME4/2 CSCs in 2012-2013.
%Small changes in materials and assembly procedures can affect the aging properties by introducing contaminants into the gas volume, so particular care was taken to adhere to the original construction procedure. 

\begin{table}[ht!]
%\begin{center}
%\large
\begin{tabular}{lc}
\hline %-------------------------------------------------------------
\hline %-------------------------------------------------------------
gas gap, $2h$ [mm]                                & 9.53   \\
wire spacing, $s$ [mm]                            & 3.13   \\
wire diameter, $d$ [$\mu$m]                       & 50   \\
wire plane-to-plane spacing, [mm]                 & 25.4   \\
strips per plane (*)                              & $\approx$10   \\
strips pitch [mm] (*)                             & $\approx$10   \\
\hline %-------------------------------------------------------------
\hline %-------------------------------------------------------------
\end{tabular}
\caption{Construction parameters of miniCSC. The parameters marked as (*) differ from the production chamber ones.}
\label{tab:miniCSCspecs}
%\end{center}
\end{table}

%%%%%%%%%%%%%%%%%%%%%%%%%%%%%%%%%%%%%%%%%%%%%%%%%%%%%%%%%%%%%%%%%%%%%%%%%%%%%%%%%%%%%%%%%%%%%%%%%%%%%%%%%%%%%%
% Description of the test stand parameters and irradiation sources used for study
\subsection{Test stand and measurement procedures} 
\label{procedure904testsetup}
\subsubsection{Irradiation}
A $^{90}$Sr source with a nominal activity of about 34 MBq was used to locally irradiate the top  chamber layer. The distance between this source and the underlying cathode plane inside the gas volume was a few~cm. The beamspot is circular with a diameter of 2.3~cm on the anode and 3.5~cm on strip planes, as estimated from the measured spatial charge distribution.
%The position of the source was marked so that the placement was consistent among irradiation sessions. 
 
 During the irradiation procedure, the top miniCSC layer ("irradiated layer") was supplied with the nominal CSC high voltage of 3600 V, while the voltage of the bottom, "reference", layer was kept off.
 The current was about 1 $\mu$A and the total wire length in the irradiated spot was estimated as 13.4~cm. The charge was accumulated at the rate of 4.3 mC/cm per day, yielding an ``acceleration factor'' of 100, corresponding to a radiation flow 100 time higher than the maximal 
maximal expected irradiation rate for the ME2/1 CSC operating under HL-LHC conditions.
The rate of the HL-LHC charge accumulation was estimated using the background hit rates and currents measured for CSC in Run 2 and corrected with FLUKA predictions for presence of a new calorimeter expected to be installed by the LH-LHC era.
%The acceleration factor of irradiation at B904 is about 3 times that used at GIF++. 
The target accumulated charge for each trial was set at 300 mC/cm, which is about 2.7 times of that expected for an ME2/1-type chamber at a 10 years-equivalent of HL-LHC operation.

For the gas system, an ``open-loop'' setup with Polyamide 11 tubing (Rilsan) was used. The flow of gas through the chamber was directed first through the reference layer and then into the irradiated one at a flow rate of 0.5 L/hour, or one layer volume per hour. Though the local irradiation conditions are difficult to scale to the uniform irradiation of CSCs at the CMS, the gas refreshment rate per a single layer was chosen to correspond to the nominal CSC flow rate of one chamber volume per six hours. No additional scaling factor was introduced to account for the accelerated irradiation.
%The reference layer was placed upstream of the irradiated layer to avoid sweeping the polymer products created in the avalanches when irradiating in the testing layer, to keep the electrodes of the reference layer clean. 
%A bubbler was connected to the output line of the irradiated layer to monitor the gas flow out of the system.

\subsubsection{Control measurements}
After each irradiation session, a set of measurements was performed in order to monitor chamber characteristics sensitive to aging effects as functions of the accumulated charge. Depending on the characteristic type, the measurements performed for the irradiated zone were compared to the ones taken for a non-irradiated part of the same layer, or for the reference layer.

Single-layer dark rates as well as the charge and rate of signals induced by an X-ray source were measured with the original CSC electronics. The dark current was measured using a Keithley 487 picoammeter. 

To monitor the stability of the gas gain, control measurements with a $^{109}$Cd source were performed in several points across the layer. Reproducibility of the source position in different measurement sessions was guaranteed by the small diameter of windows in the upper chamber panel. The standard CSC electronics allows read out of the matched discriminated signals from the cathode strips and anode wire groups and, if such a coincidence occurs, the charge induced on the strips. The charge is shared among three neighbour strips forming a cluster. The measurements were done for irradiated and the control non-irradiated points of the same layer, and the ratio of those values represents the relative gas gain free of the pressure and temperature fluctuations. The rate of the signals was measured as a function of the high voltage and provides the additional information on the gas gain stability and also on the presence of after-pulses or sparks related to the avalanche. The rate measurements were done for the whole layer and were compared to the rate measured when no source is present, i.e. dark rate.

The dark rates and dark currents were measured for the voltage range of 2.5-3.8 kV. The dark current was measured between the joined strips and ground. Each measurement was performed for the period 10--15 min, and the average value of the current after settling was taken as the result. To control the humidity influence, the corresponding current measurements were done for the non-irradiated layer during the same day.

%\subsubsubsection{Cathode-Related Observables} %\label{procedure904cathodeMeas}

% Monitoring of Malter effect/currents
Cathode aging may appear as a deposition of a thin film which can increase conductivity across adjacent cathode strips causing degradation of the spatial resolution of the chamber. The strip-to-strip resistance was monitored  measuring current between two adjacent strips when 300~V was applied to one of them and second was grounded.

If cathode deposits are insulating and thin, and the chamber occupancy is relatively high, the effect of Malter currents may appear. This type of current is caused by emission from the cathode surface when a positive charge on the insulating film is high enough, and may be persistent even when the irradiation is stopped. During the irradiation, the current was monitored and checked for unstable behaviour and spikes. Additionally, every time the $^{90}$Sr source was removed from the chamber, the dark current was monitored for 10-15 minutes.

\subsection{Additional studies}
\label{sec:gifprocedures}
%\fixme{ Rediced to miniCSC - here it's a mixture of large chamber and the miniCSC at GIF++ which was open loop in fact}\\

To cross-check the local irradiation test results, an additional miniCSC was irradiated at GIF++ with 2\% CF$_4$ gas mixture. GIF++ provides nearly uniform radiation over the whole chamber, which is more representative of the exposure in CMS. The irradiation acceleration factor of this test was of about 30. The control measurements procedure for this prototype was the same as in the case of local irradiation.

Additionally a local irradiation test with a CSC prototype was performed at PNPI for the nominal CSC gas mixture. The accumulated charged collected during the test was 1.2 C/cm. No change in the prototype detection performance was observed~\cite{Buzoveria:2020itl}.

%% file: LongevityResults.tex
\section{Longevity Results with varying CF$_{4}$ fractions} 

The irradiation and measurement procedures described in Section \ref{procedure904} were carried out on 
miniCSCs with three different gas mixtures of 
Ar/CO$_{2}$/CF$_{4}$: 40\%/55\%/5\%, 40\%/58\%/2\%, and 40\%/60\%/0\%.
It should be noted that the 2\% CF$_{4}$ trial was performed after the 5\% CF$_{4}$ one using the same chamber layer. For the 2\% CF$_{4}$ test the irradiation zone was displaced to a distant point of the layer, and the small size of the irradiated area ensured that the influence of the previous exposure was negligible. The reference layer was the same in both trials. For the 0\% CF$_{4}$ study a different miniCSC was used. No significant change in the characteristics of the miniCSC under test was observed in any of the tests. Similar results were obtained in the additional uniform irradiation test with 2\% CF$_{4}$ mixture at GIF++. 

\subsection{Gas gain} 
\begin{figure}[htbp]
\begin{center}
\includegraphics[height=0.28\textwidth]{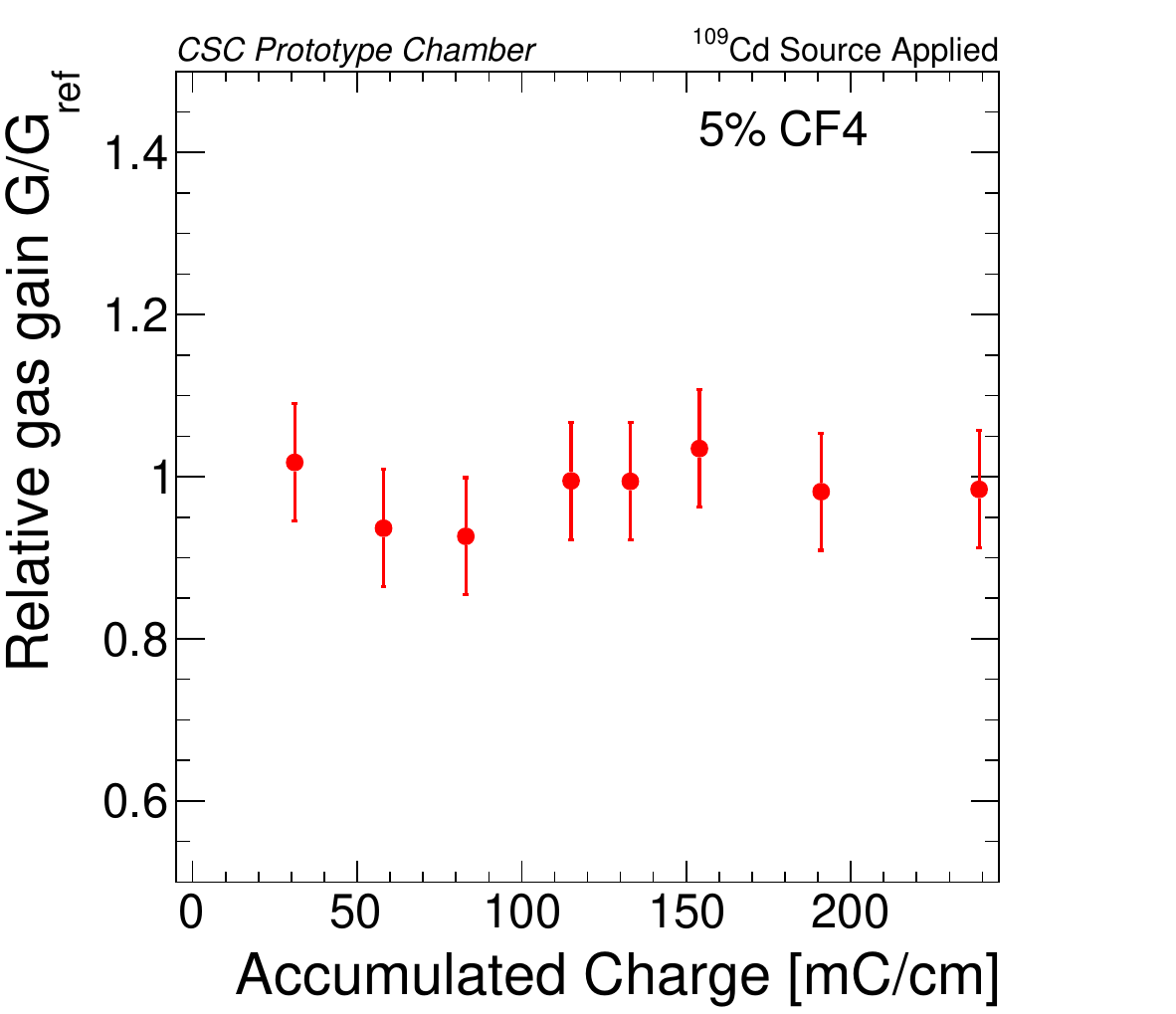}
\includegraphics[height=0.28\textwidth]{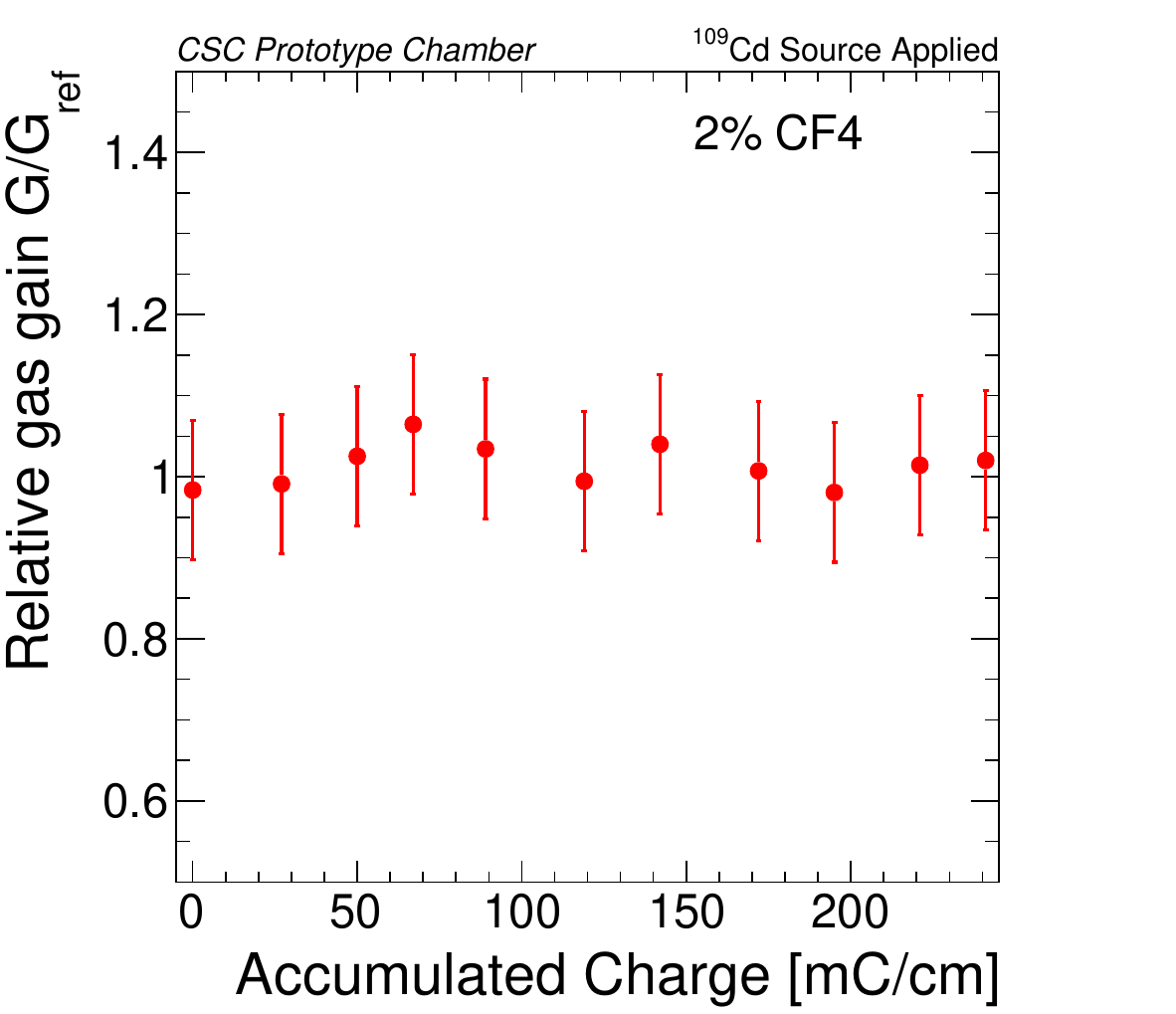}
\includegraphics[height=0.28\textwidth]{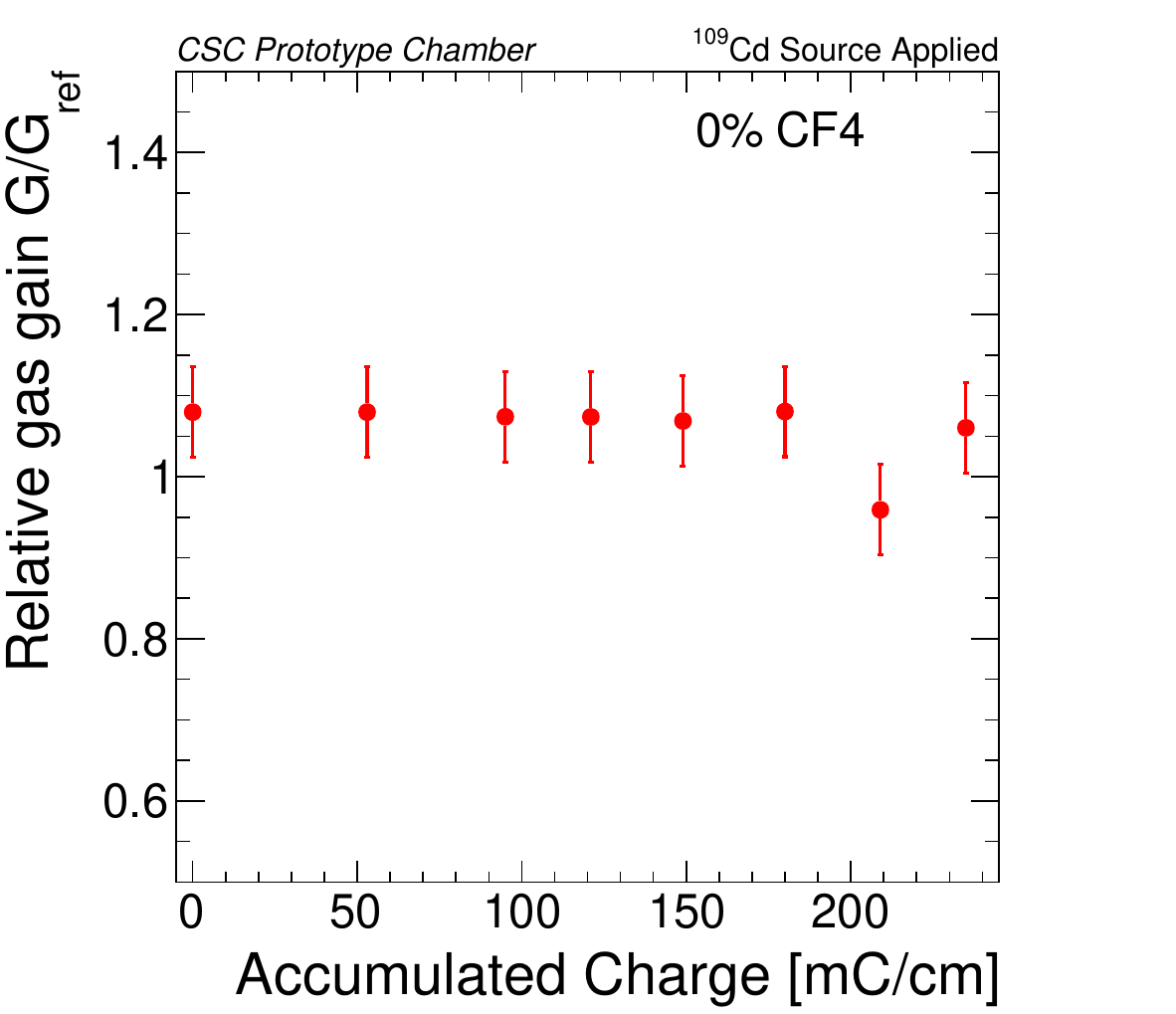}
\caption{Relative gas gain as the peak position of a $^{109}$Cd X-ray spectrum measured in the irradiated area and normalized to that of non-irradiated reference zones for gas mixtures with 5, 2, and 0\% CF$_{4}$ (left, center, right).
    \label{fig:relativeGG}}
\end{center}
\end{figure}

Figure~\ref{fig:relativeGG} shows changes in the relative gas gain obtained as a ratio of the cluster charge peak position measured for the irradiated and  control areas, for gas mixtures with 5, 2, and 0\% CF$_{4}$. The error bars represent statistical uncertainties. Stable behaviour was observed during the whole irradiation period up to 240~mC/cm.

\subsection{Dark rate and Cd rate measurements} 

% Dark Rates ========

\begin{figure}[htbp]
\begin{center}
\includegraphics[height=0.28\textwidth]{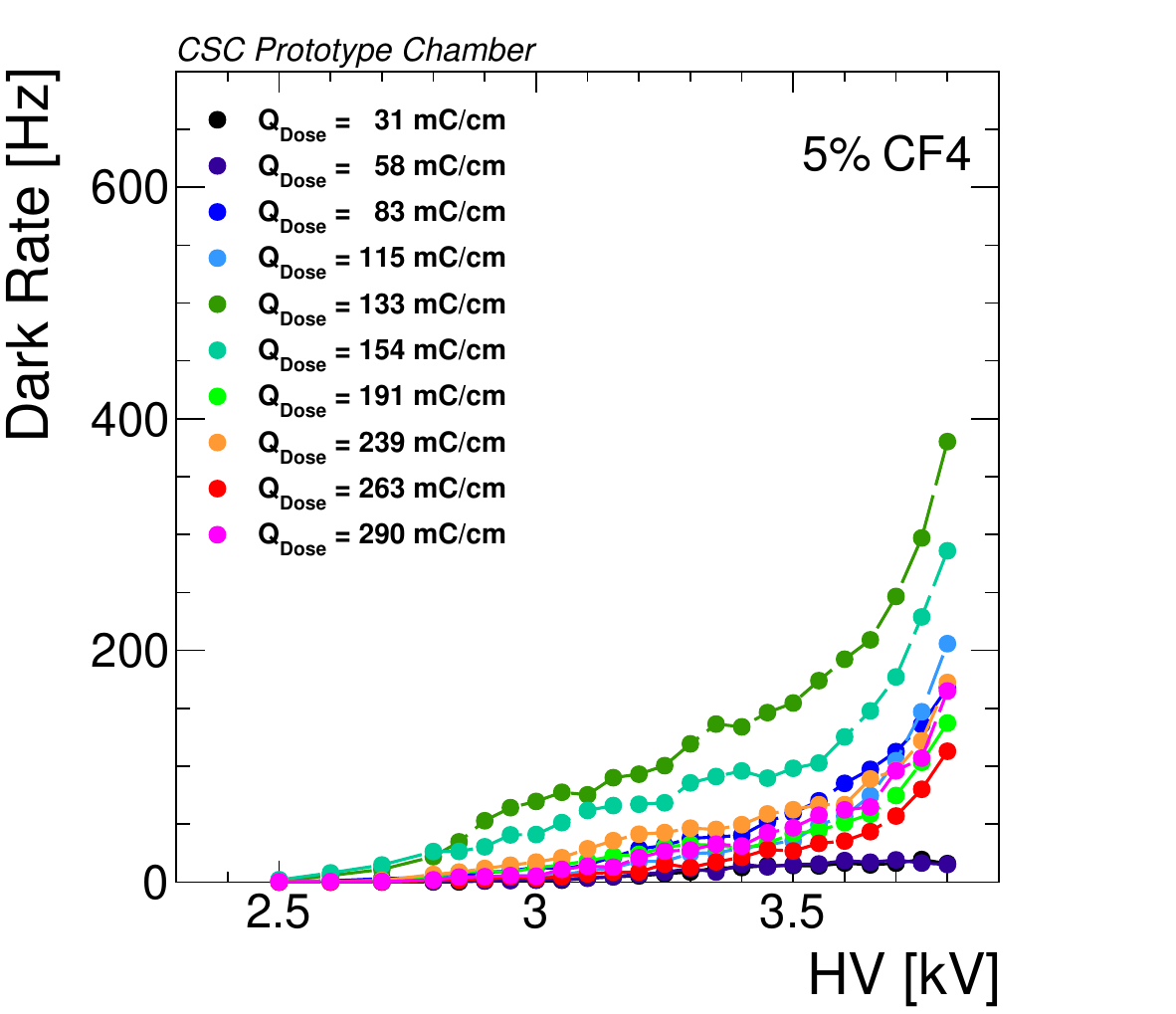}
\includegraphics[height=0.28\textwidth]{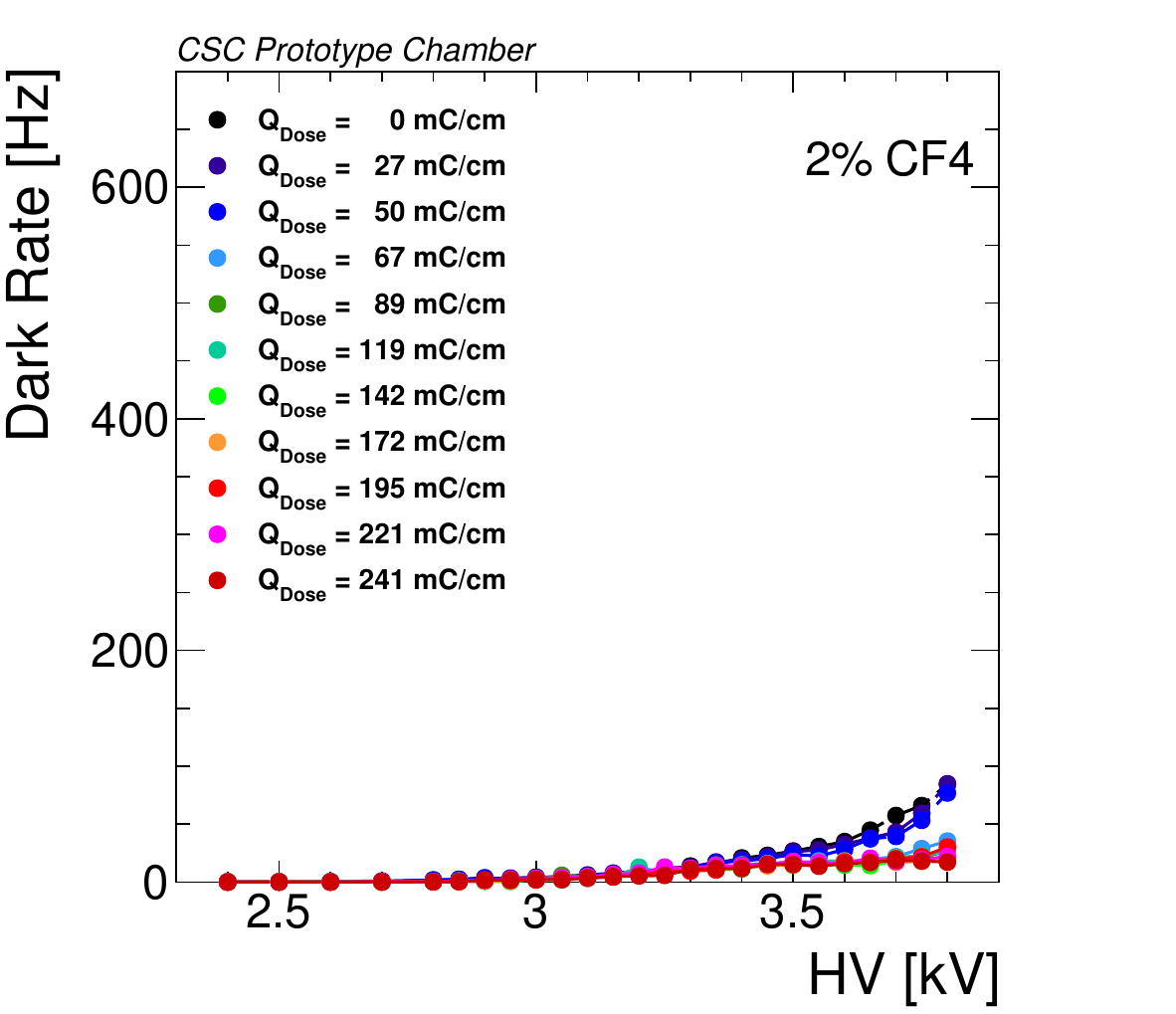}
\includegraphics[height=0.28\textwidth]{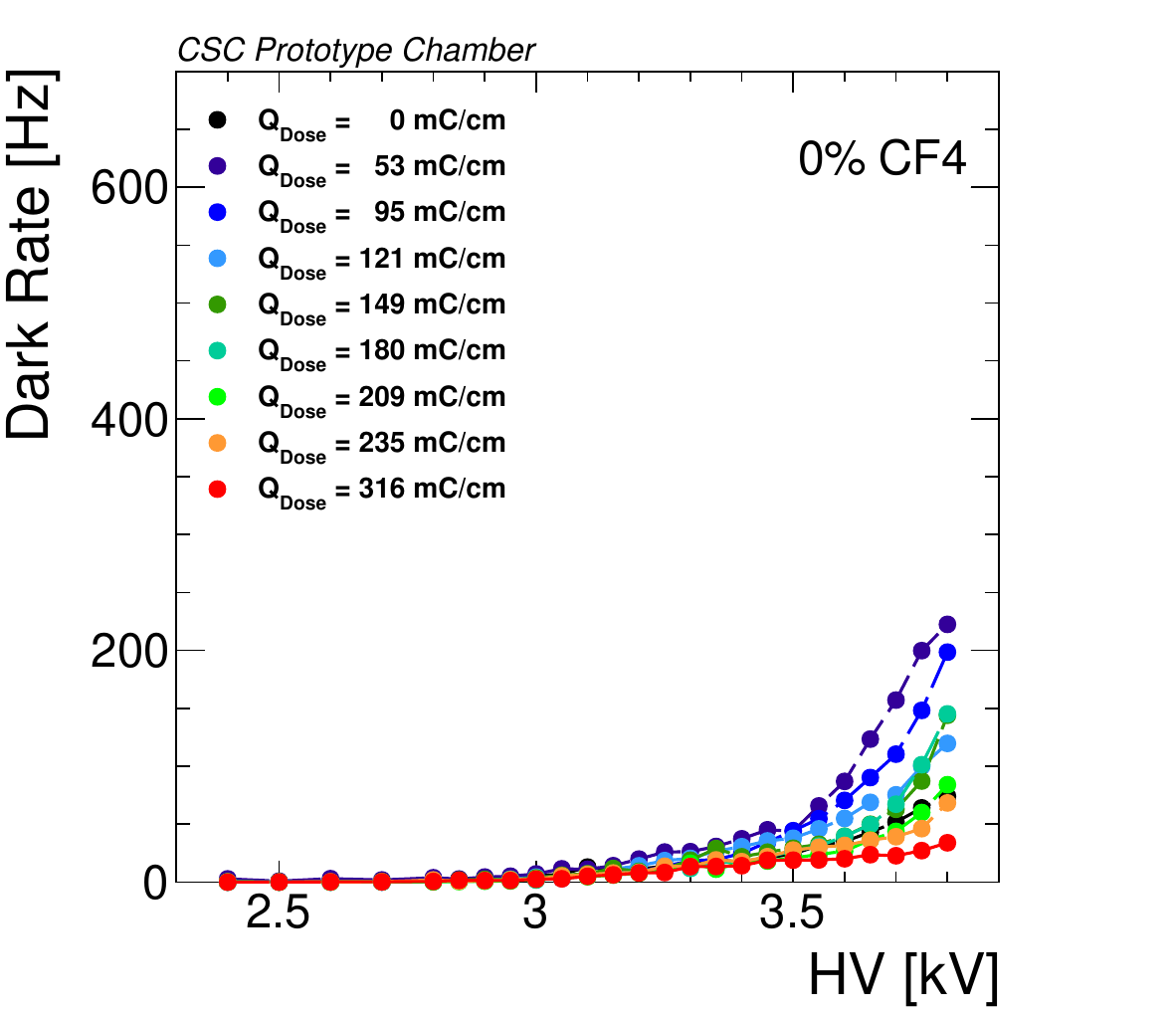}
\caption{Dark rates for gas mixtures with 5, 2, 0\% CF$_{4}$ (left, center, right) as a function of high voltage measured for different accumulated charge $Q$. 
%{\color{red}(Should be that rates were specifically collected for the irradiated hole in each of the different trials)}
    \label{fig:darkratesHV}}
\end{center}
\end{figure}

Figure~\ref{fig:darkratesHV} shows the dark rate per miniCSC layer as a function of the high voltage  measured for different values of the accumulated charge $Q$ for gas mixtures with 5, 2, and 0\% CF$_{4}$. There is an overall decrease in dark rates in the 2\% CF$_{4}$ trial compared to the 5\% trial.  This is possibly an effect of the ``HV training'' of the chamber layer, since this same layer was used in both trials. 
The differences in dark rate are insignificant, both as a function of the accumulated irradiation dose and gas mixture type. 
Thus we conclude that the dark rate is stable with respect to exposure for all three gas mixtures.

% Source rates only (dark rates subtracted) -------
%\subsubsection{Count Rates with Background Subtraction}

\begin{figure}[h]
\begin{center}
\includegraphics[height=0.28\textwidth]{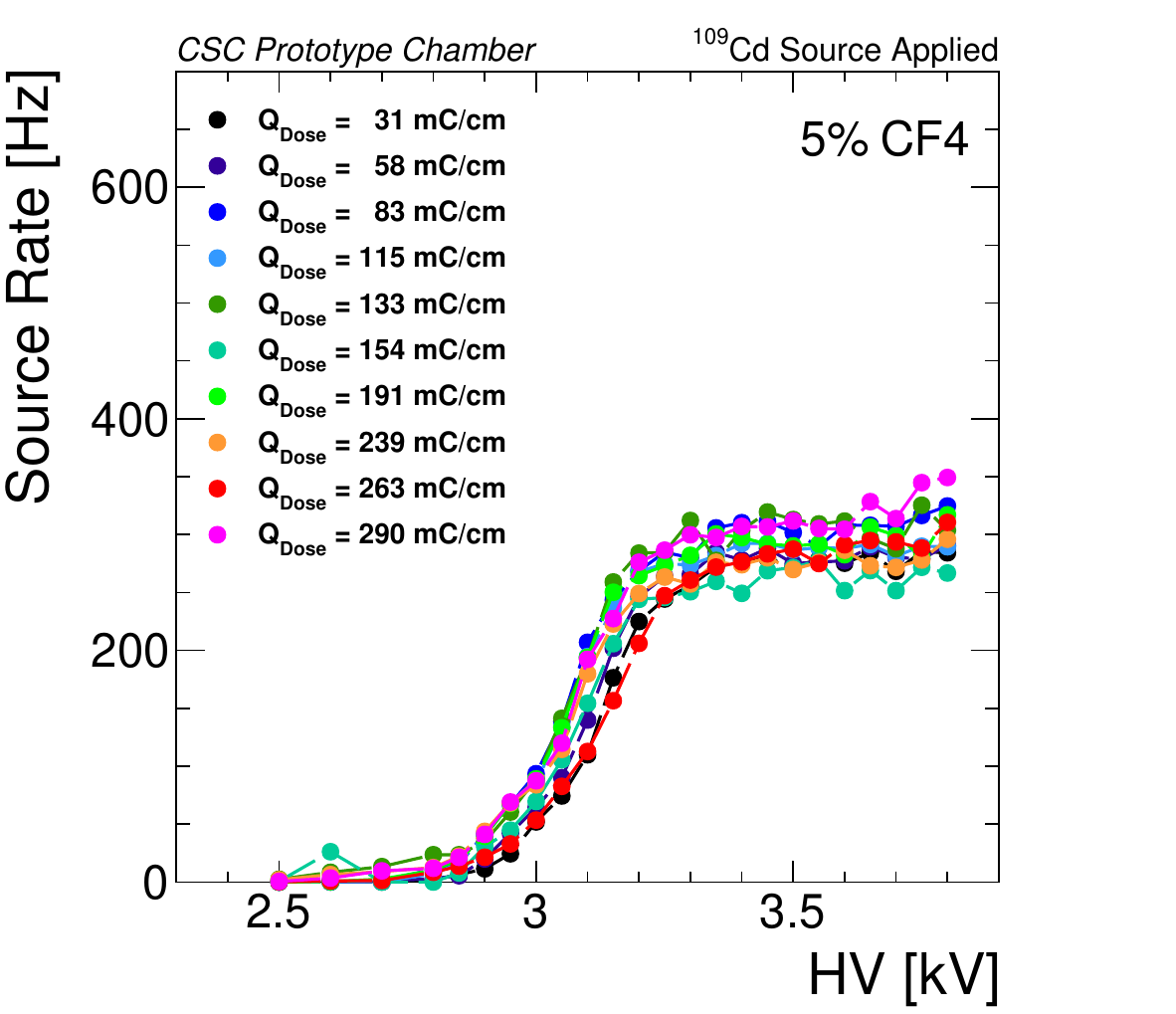}
\includegraphics[height=0.28\textwidth]{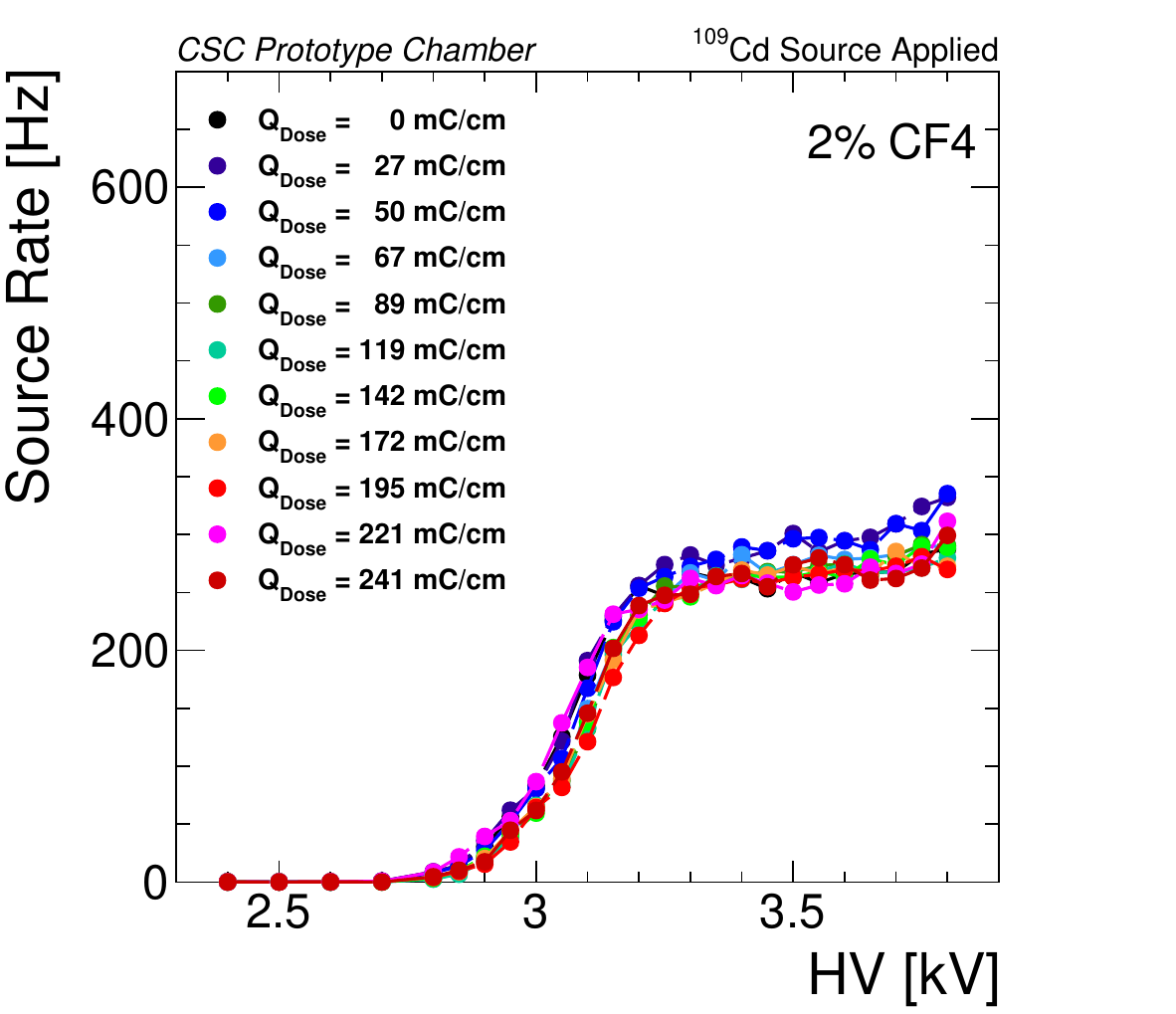}
\includegraphics[height=0.28\textwidth]{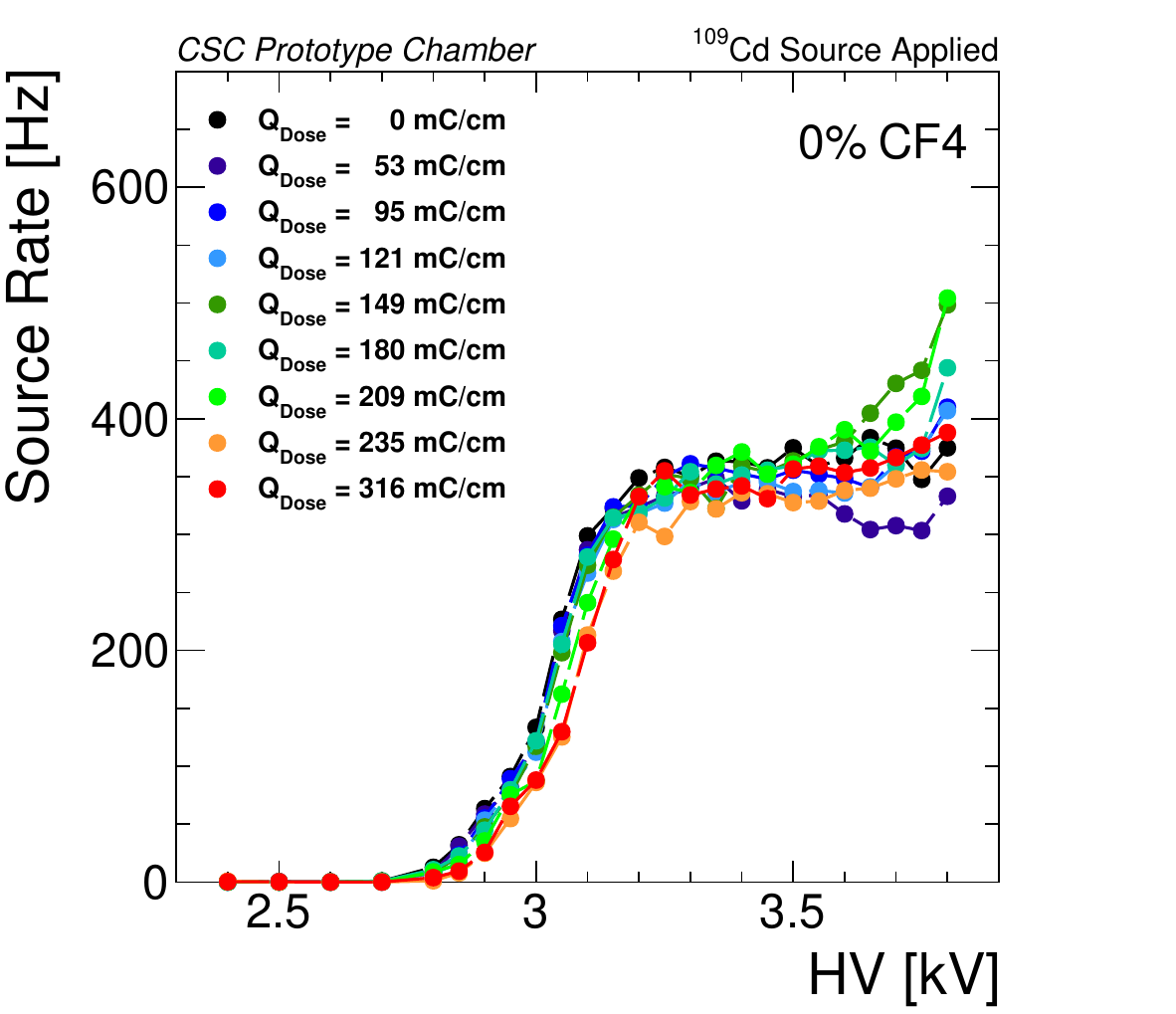}
\caption{Single-layer count rates for gas mixtures with 5, 2, and 0\% CF$_{4}$ (left, center, right) obtained with a $^{109}$Cd X-ray source as functions of the high voltage for different accumulated charges. The measurements are corrected for the dark rate contribution and for the decrease of $^{109}$Cd source intensity over the period of the three tests.
    \label{fig:sourceratesHV}}
\end{center}
\end{figure}

\begin{figure}[h]
\begin{center}
\includegraphics[height=0.28\textwidth]{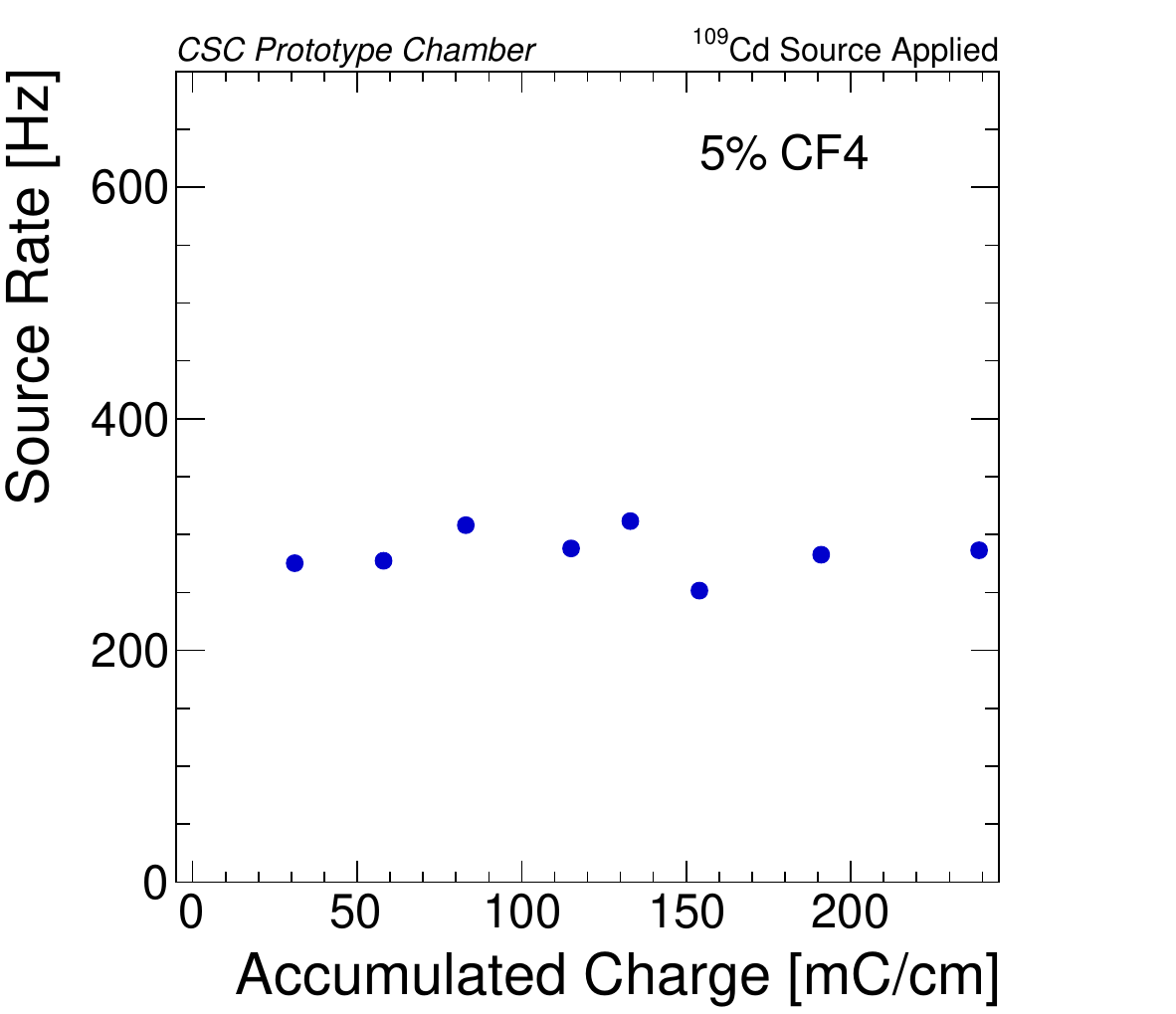}
\includegraphics[height=0.28\textwidth]{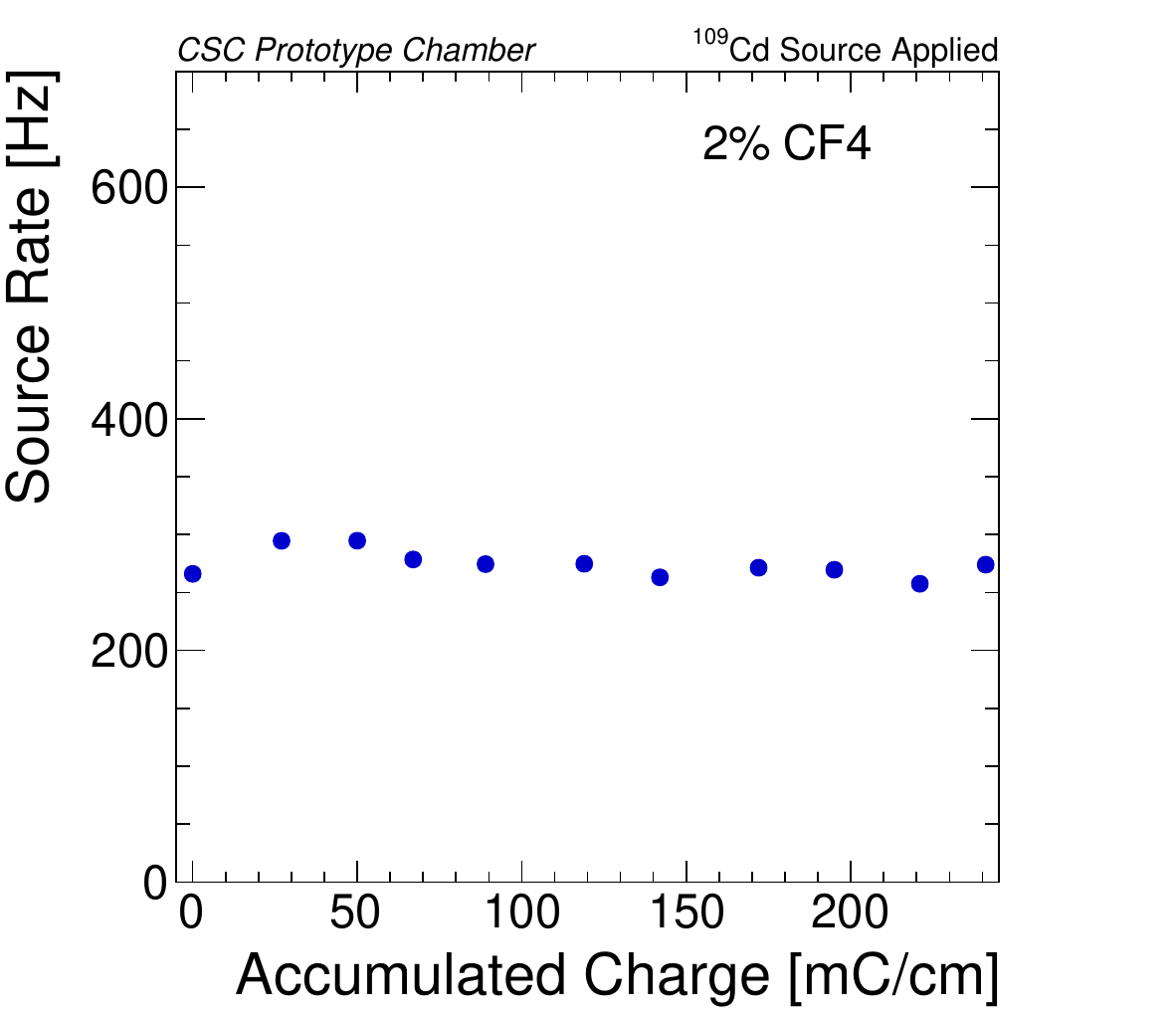}
\includegraphics[height=0.28\textwidth]{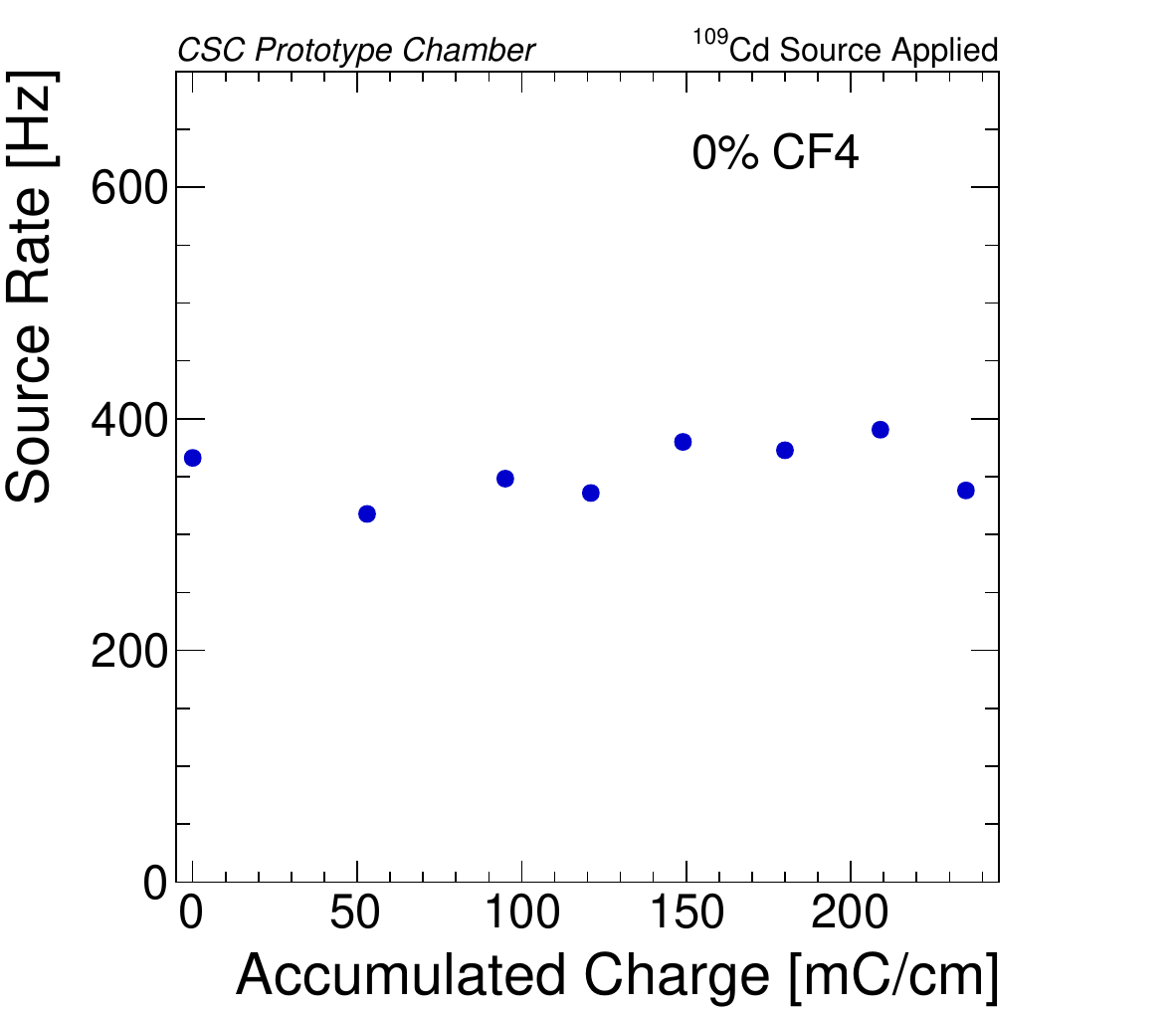}
\caption{Single-layer count rates for gas mixtures with 5, 2, 0\% CF$_{4}$ (left, center, right) obtained with a $^{109}$Cd X-ray source for the operating point of 3.6 kV as functions of the accumulated charge.
    \label{fig:sourceratesWP}}
\end{center}
\end{figure}

Figure~\ref{fig:sourceratesHV} shows the source rates as a function of the high voltage, collected for gas mixtures with 5, 2, and 0\% CF$_{4}$, and monitored as a function of the accumulated charge $Q$. The contribution from dark rates, measured separately and as shown in Fig.~\ref{fig:darkratesHV}, was subtracted. Then the remaining source rates were corrected for the decrease in the $^{109}$Cd source radioactivity. The shape of the efficiency curve and the value of the plateau are very similar for the three cases. The slight increase in the count for the 0\% CF$_{4}$ measurements can be explained by  a smaller thickness of the window for positioning the $^{109}$Cd source. No significant change in the efficiency dependence was observed during irradiation for all three gas mixtures.
Figure~\ref{fig:sourceratesWP} shows the source count rate at the operating point of 3.6 kV, as a function of the accumulated charge. 

%\begin{figure}[h]
%\begin{center}
%\includegraphics[height=0.28\textwidth]{images/LongevityResults_904/5percent/5percent_DarkCurrent_3p6.pdf}
%\includegraphics[height=0.28\textwidth]{images/LongevityResults_904/2percent/2percent_DarkCurrent_3p6.pdf}
%\includegraphics[height=0.28\textwidth]{images/LongevityResults_904/0percent/0percent_DarkCurrent_3p6.pdf}
%\caption{Dark current values for both the irradiated and reference layers, taken at the working point voltage, for 5, 2, 0\% CF$_{4}$.
%    \label{fig:darkcurrentWP}}
%\end{center}
%\end{figure}

\subsection{Dark current and strip-to-strip resistance}
% Dark Current ======
The dark current was measured using the procedure outlined in Section \ref{procedure904}. The values at the working point were of order 10$^{2}$ to 10$^{3}$ pA for all trials and mostly defined by the leakage currents. Neither current spikes nor dependence on the accumulated charge was observed. %By comparison, values of the current when operating the chamber at the working point voltage (3.6 kV) when using the $^{90}$Sr source for irradiation were ~1 microAmpere. %{\color{red}(Need to add a conclusion statement.)}

% {\color{red}(For 2\% and 0\% CF$_{4}$ trials, put in the dark current HV scan plots for irradiated and reference layers like those depicted in Fig.~\ref{fig:darkcurrent_GIFminiChamber}, center and right plots. Make sure to also describe behavior of these HV scans plots to be added.)}

\begin{figure}[htbp]
\begin{center}
\includegraphics[height=0.31\textwidth]{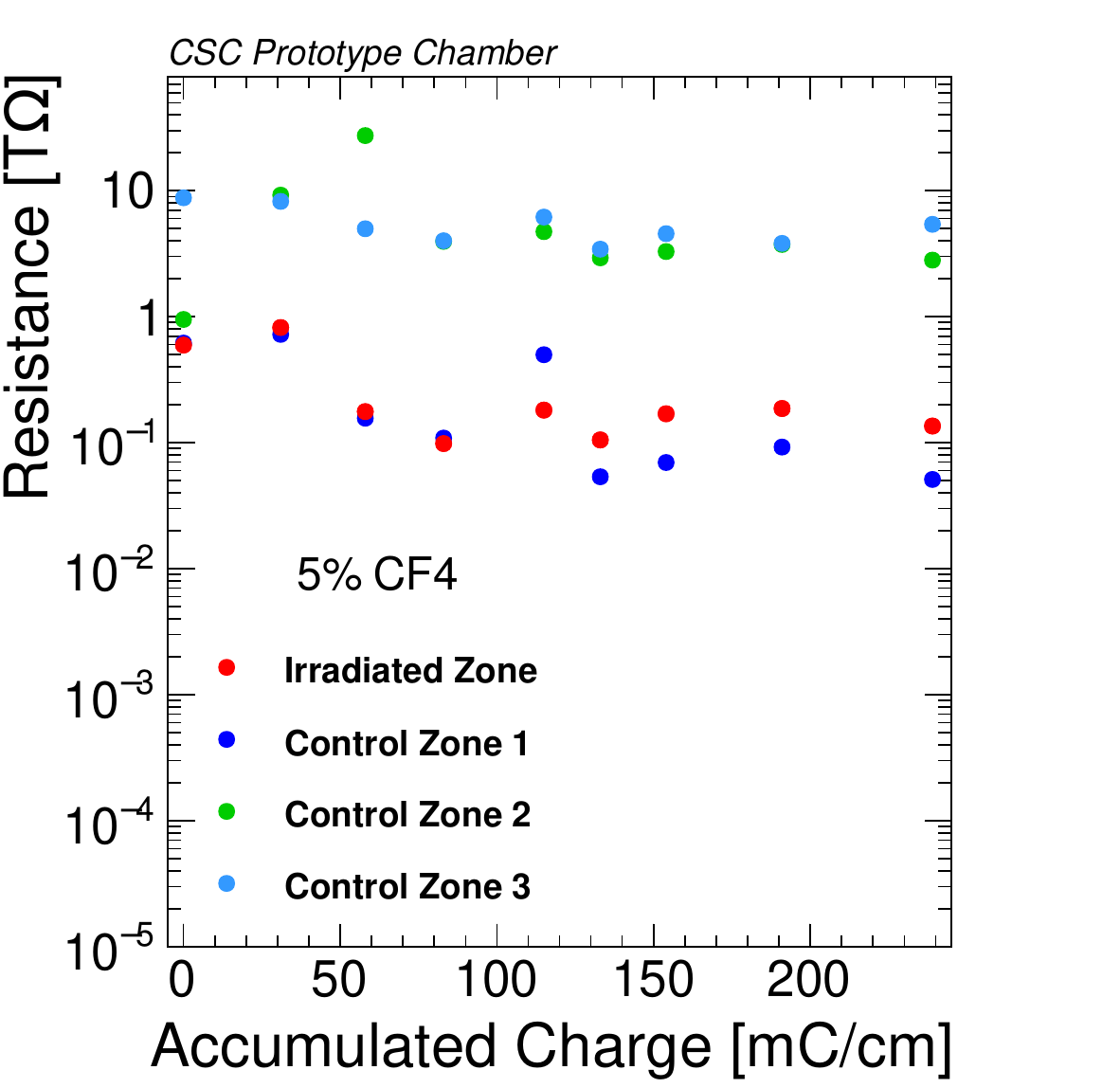}
\includegraphics[height=0.31\textwidth]{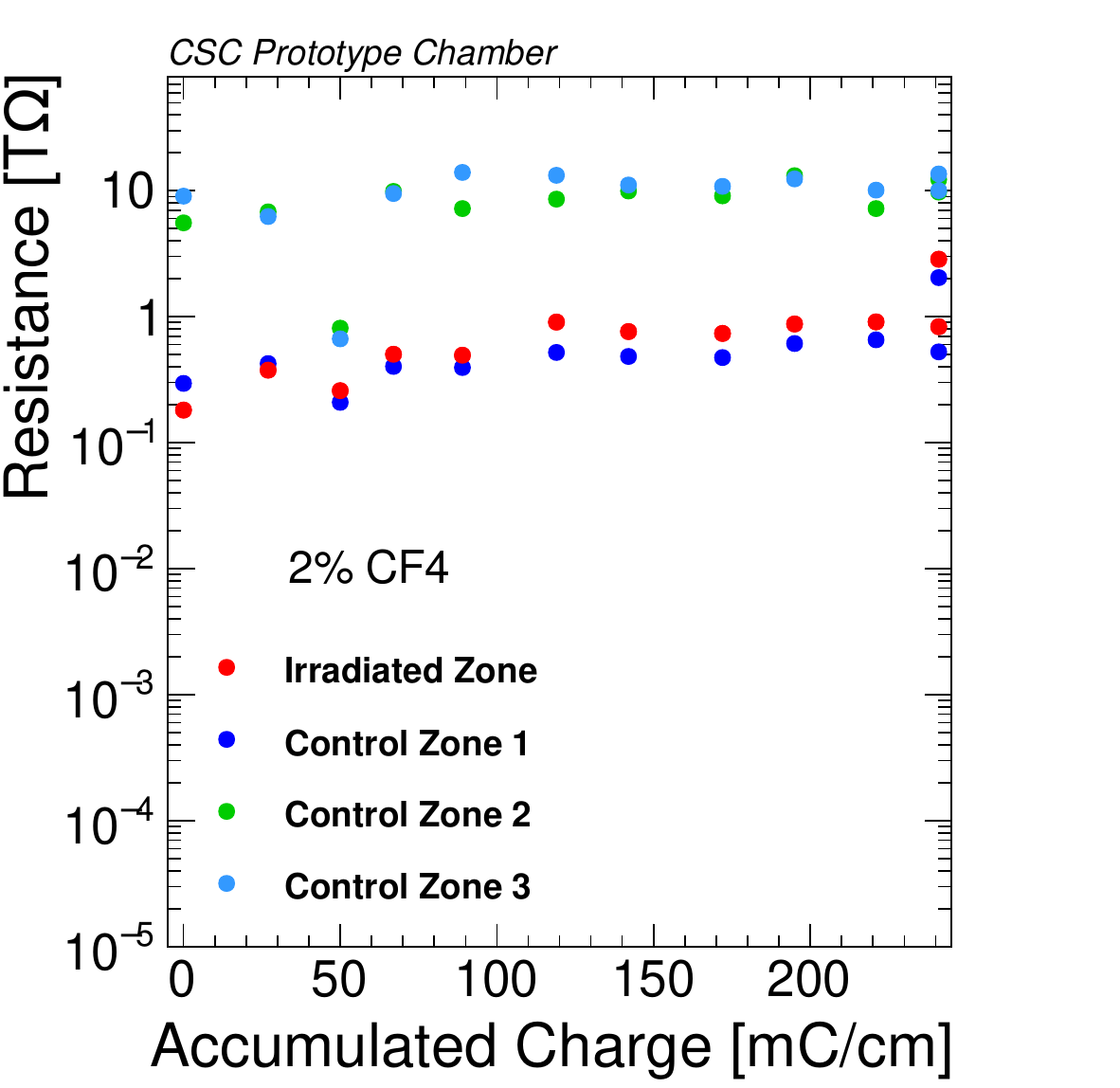}
\includegraphics[height=0.31\textwidth]{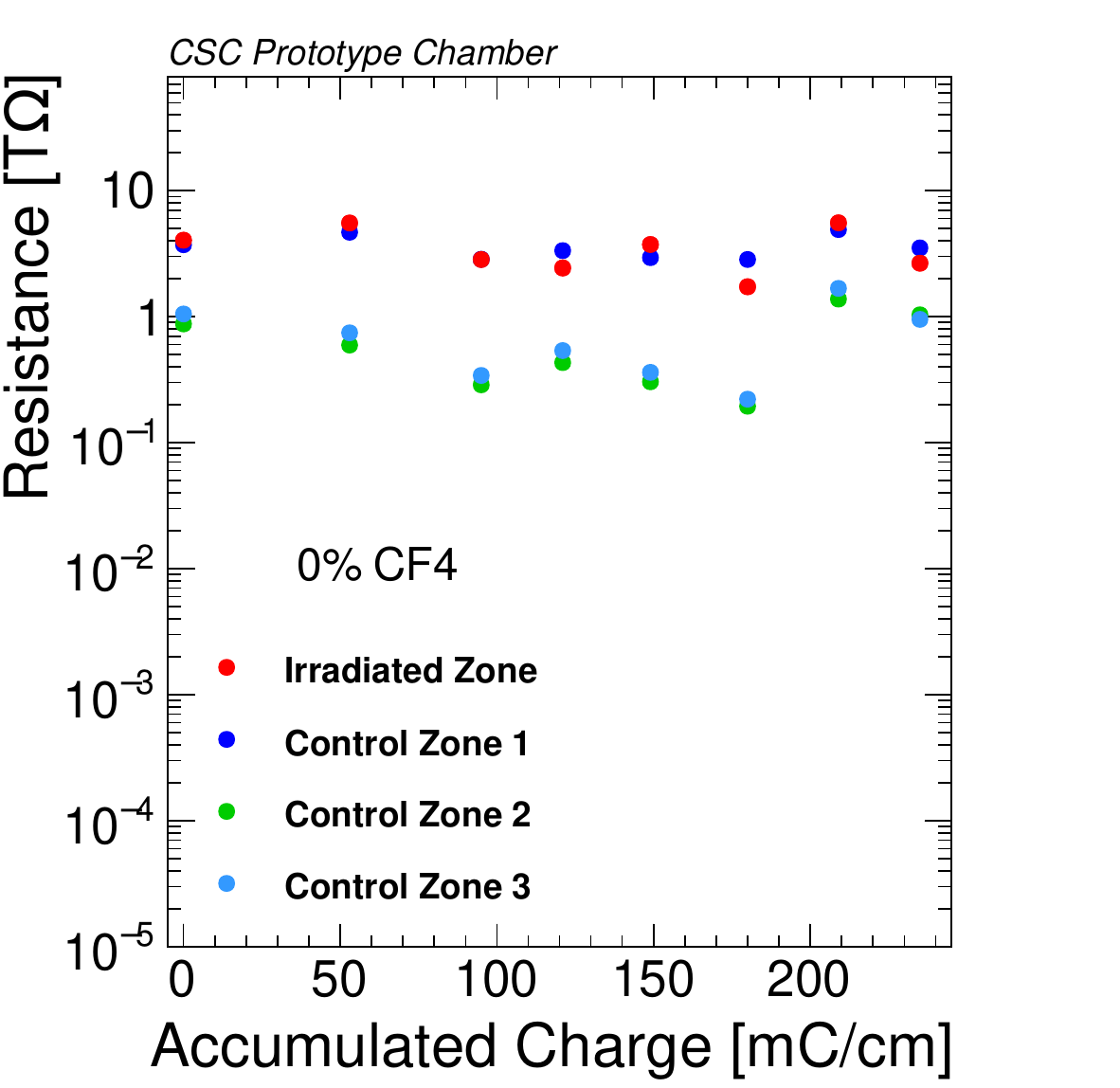}
\caption{Electrical resistance as measured between pairs of adjacent cathode strips on a single chamber layer, shown for gas mixtures with 5, 2, and 0\% CF$_{4}$ (left, center, right). Strips are selected in the irradiated area (red), outside the irradiated area but at the same layer (blue) and at the reference layer (ligh blue and green). The systematic difference in the measured resistances for different layers are defined by the CSC construction.
    \label{fig:StS}}
\end{center}
\end{figure}

Figure~\ref{fig:StS} shows electrical resistance measured between pairs of adjacent cathode strips, each on a single chamber layer, shown for gas mixtures with 5, 2, and 0\% CF$_{4}$.

%%%%%%%%%%%%%%%%%%%%%%%%%%%%%%%%%%%%%%%%%%%%%%%%%%%%%%%%%%%%%%%%%%%%%%%%%%%%%%%%%%%%%%%%%%%%%%%%%%%%%%%%%%%%%%

%\input{LongevityResults_GIF++}

%\input{LongevityResults_PNPI}

%% file: AnalysisOfElectrodes.tex
\section{Analysis of Irradiated Electrodes} 
\label{analysisElectrodes}

After irradiation up to about 300 mC/cm, the miniCSCs were disassembled and visually inspected. Then, samples of anode wires and cathode planes, both from the irradiated and non-irradiated zones, were prepared for optical and spectroscopic analyses, as well as virgin samples of anode wires and cathode planes. Preliminary analysis was performed at CERN MM-MME department, while final measurements were done at the Physical Chemistry Faculty of the Belgrade University. 
%{\color{gray}
%For the case of the 5\% and 2\% CF$_{4}$ trials, the chamber was opened after both trials were completed. First, photographs of the cathodes and anodes were taken for visual inspection. Afterwards, samples of the anode wires and the copper cathode layer were cut out and examined using several spectroscopic techniques, as described in Section \ref{analysisElectrodesBelgrade}. 
%}
%%%%%%%%%%%%%%%%%%%%%%%%%%%%%%%%%%%%%%%%%%%%%%%%%%%%%%%%%%%%%%%%%%%%%%%%%%%%%%%%%%%%%%%%%%%%%%%%%%%%%%%%%%%%%%
% Photos of irradiated cathodes, anodes
\subsection{Optical inspection of the irradiated electrodes}
\begin{figure}[ht]
\begin{center}
\includegraphics[height=0.25\textwidth]{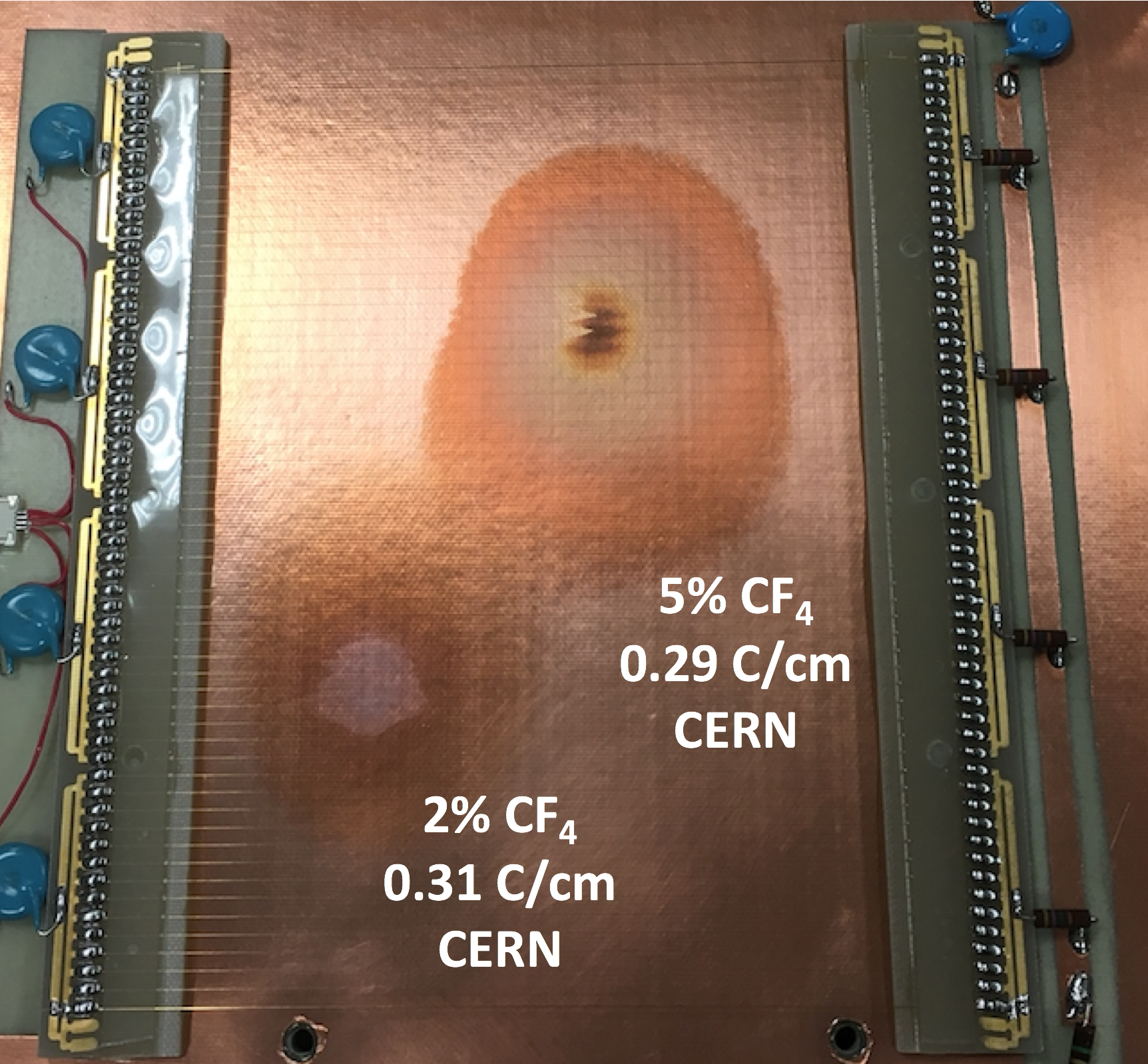}
\hspace{0.4cm}
\includegraphics[height=0.25\textwidth]{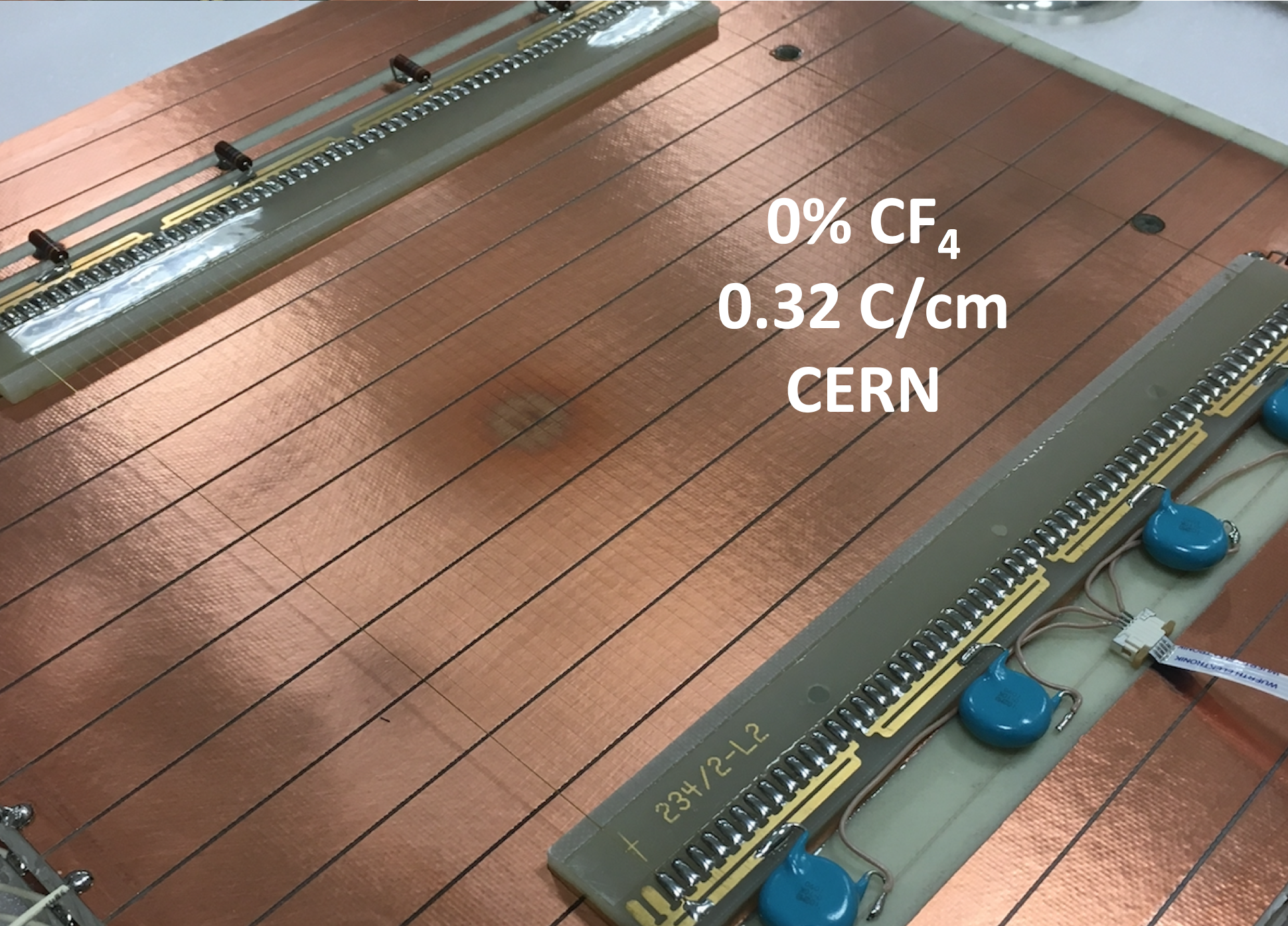}
\caption{Photos of the irradiated layer for the miniCSC chambers used in the 5, 2, 0\% CF$_{4}$ trials, with the total accumulated charge indicated. Effects of irradiation can be easily seen on the cathode.
    \label{fig:exposedCathodePhotos}}
\end{center}
\end{figure}

Photos of the exposed cathodes, taken just after opening the irradiated layer of the miniCSC, for the 5, 2, and 0\% CF$_{4}$ trials are shown in Fig.~\ref{fig:exposedCathodePhotos}. In all cases, there is a circular spot that corresponds to the primary area of irradiation where the $^{90}$Sr was placed on top of the layer. Also, a circular area around the irradiated spot is showing a radial color-gradient feature. The size of the spot decreased as the CF$_{4}$ fraction in the gas mixture decreased. The most important observation was visibly darkened anode wires in the areas irradiated with 2\% CF$_{4}$ and 0\% CF$_{4}$ while no such effect was noticed for the 5\% CF$_{4}$ irradiation zone. 

\begin{figure}[h]
\begin{center}
\includegraphics[width=0.6\textwidth]{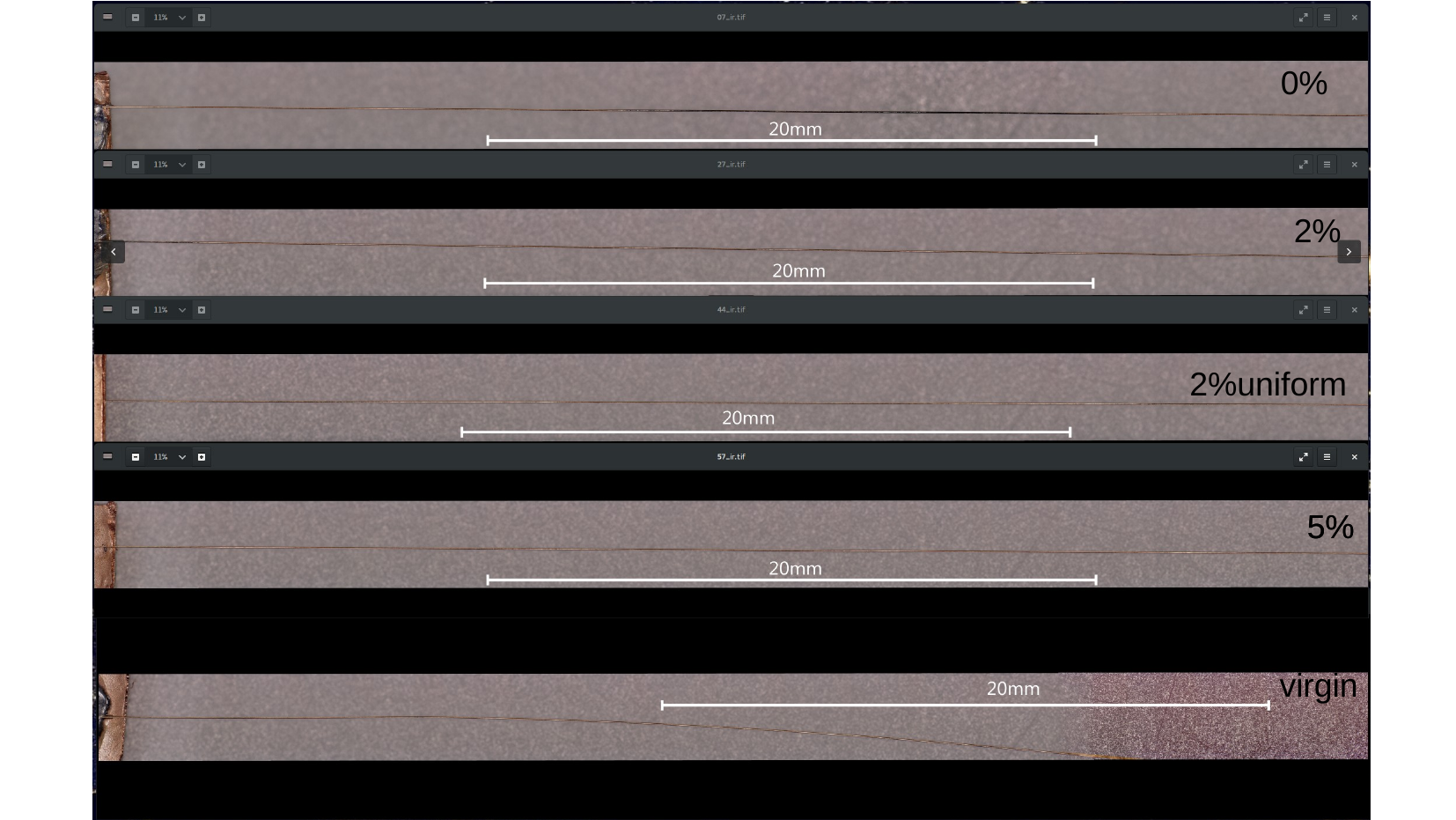}
\caption{Optical microscopy snapshots for anode wires irradiated with 5, 2, 0\% CF$_{4}$ gas mixtures compared to a virgin sample.
    \label{fig:wiresopt}}
\end{center}
\end{figure}

Figure~\ref{fig:wiresopt} shows snapshots of the optical microscope images comparing anode wires irradiated with 5, 2, 0\% CF$_{4}$ gas mixtures and a virgin sample. The darkening is clearly seen for the 0\% CF$_{4}$ sample and for the 2\% CF$_{4}$ locally irradiated sample. Wire darkening can be noticed also for 2\% CF$_{4}$ uniform irradiation, while no color change is seen for the 5\% CF$_{4}$ sample compared to the virgin wire.

\subsection{Elemental analysis of the electrodes}
\begin{figure}[h]
\begin{center}
\includegraphics[width=0.6\textwidth]{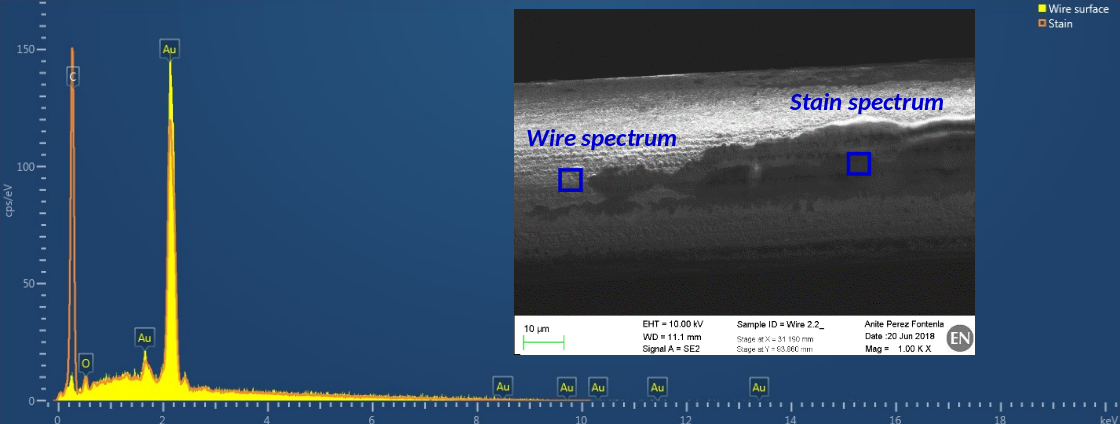}
\caption{SEM image and the EDS spectra of different areas of the 2\% CF$_{4}$ anode wire sample.
    \label{fig:wireCERN}}
\end{center}
\end{figure}
\begin{figure}[htbp]
\begin{center}
\includegraphics[width=0.31\textwidth]{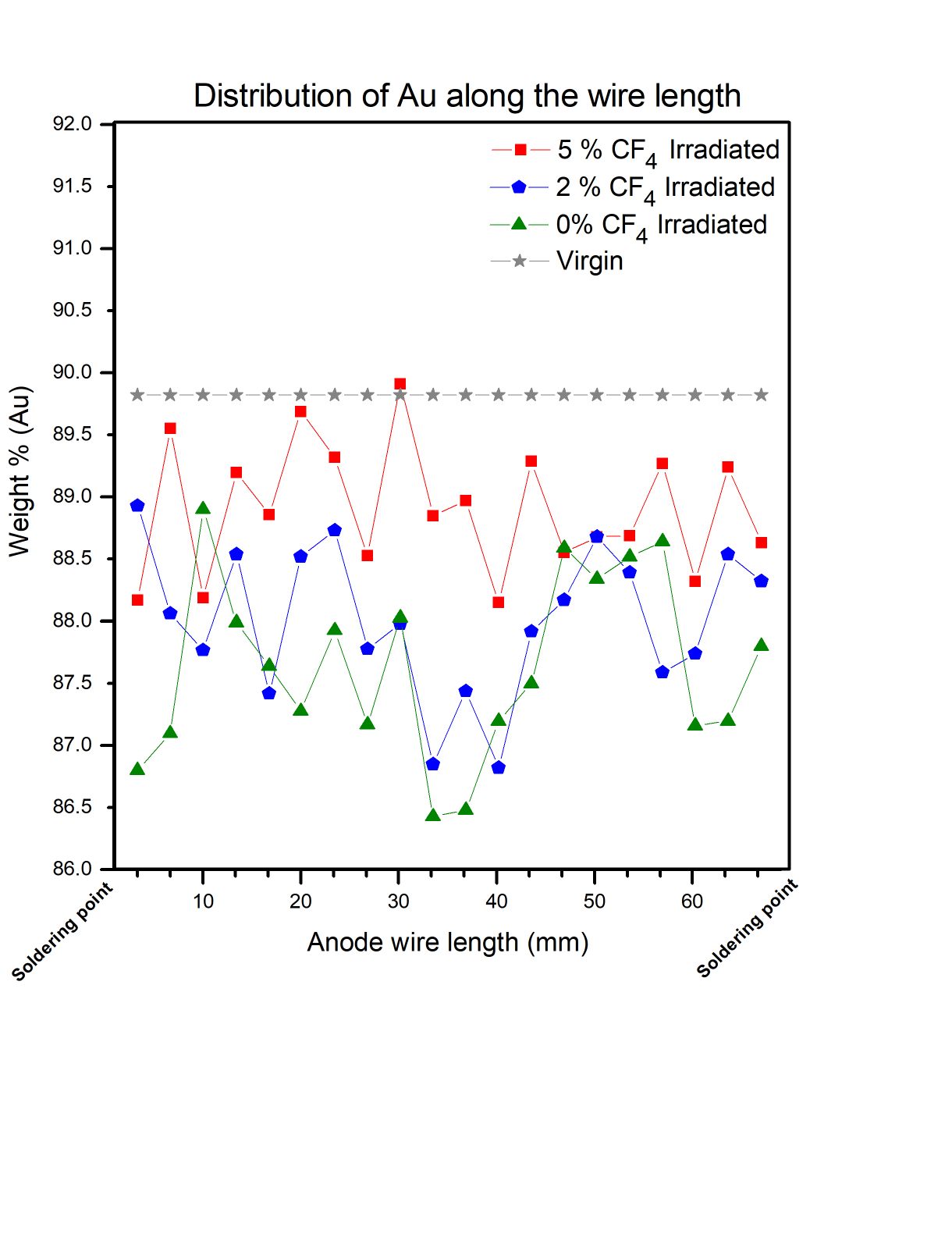}
\hspace{0.1cm}
\includegraphics[width=0.31\textwidth]{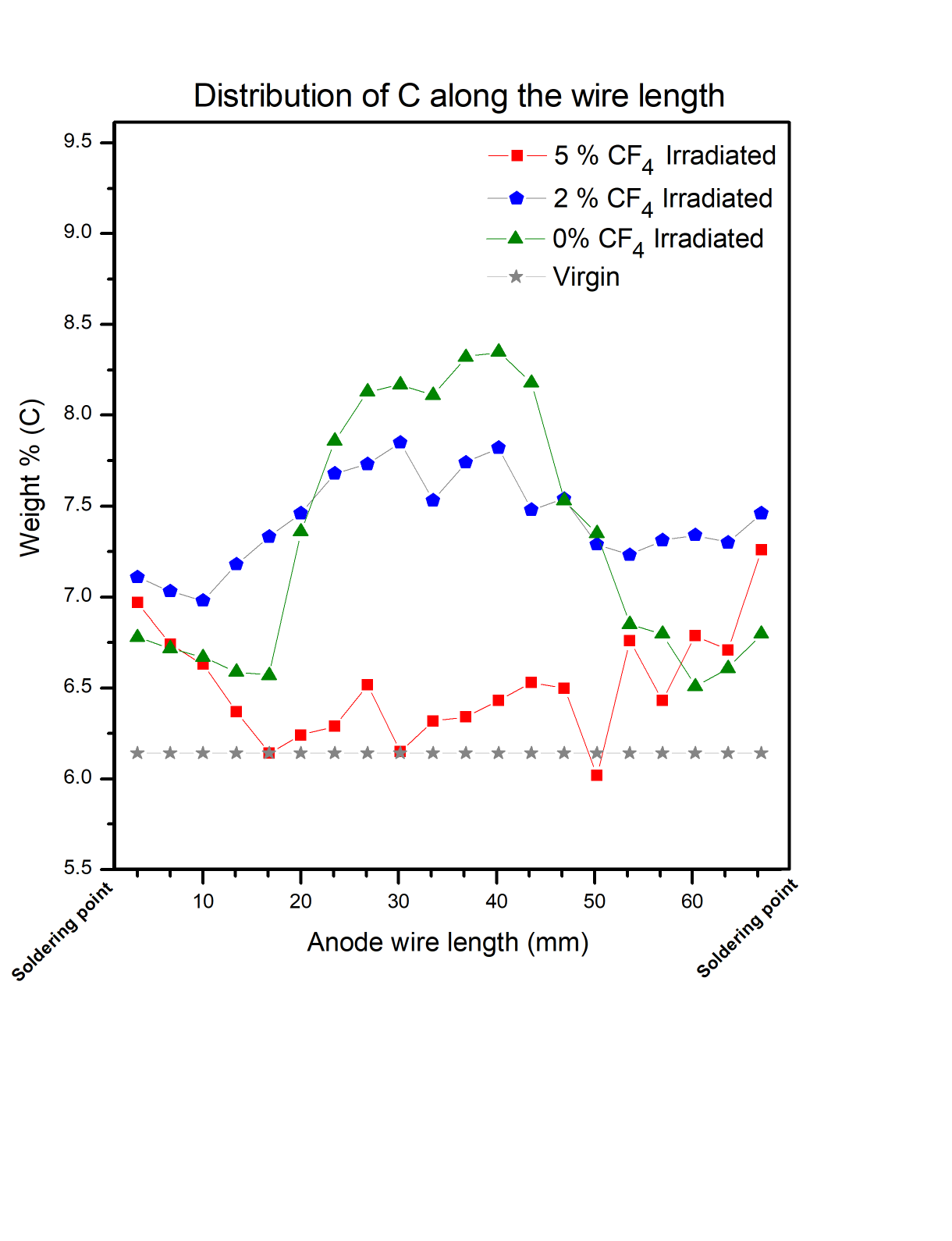}
\hspace{0.1cm}
\includegraphics[width=0.31\textwidth]{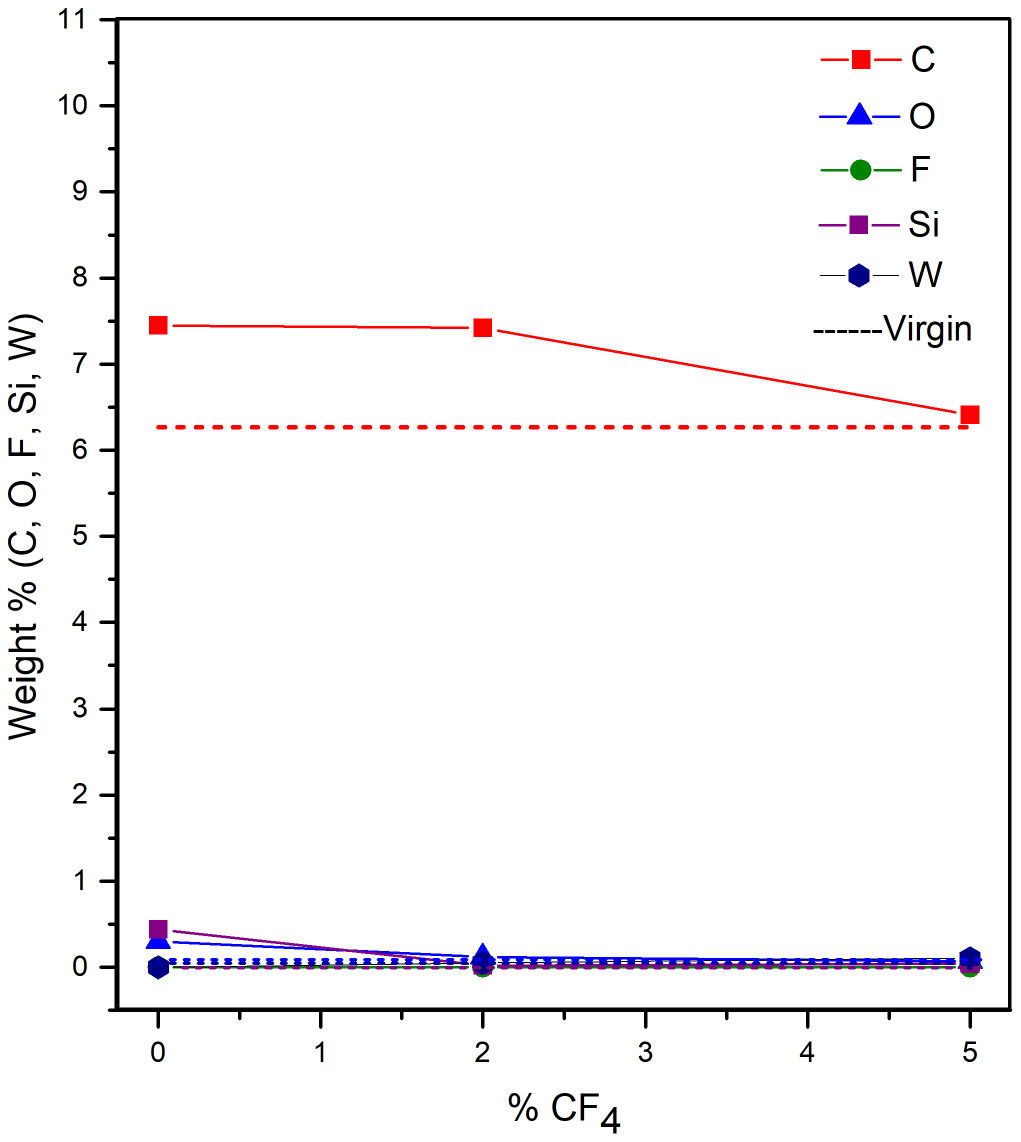}
\hspace{0.1cm}
\caption{EDS measurement results for anode wires: relative weight fractions for gold (left) and carbon (middle) along the wire length and the average values for carbon, silicon, oxygen, fluorine and tungsten as functions of the CF$_{4}$ fraction in the gas mixture. 
    \label{fig:EDSwires}}
\end{center}
\end{figure}

Morphological and micro-elemental analysis of the electrodes was done using 
Scanning Electron Microscopy (SEM) along with Energy Dispersive Spectrometry (EDS). Preliminary analysis of the anode wires showed presence of carbon for 2\% and 0\% CF$_{4}$ samples, as can be seen in Fig.~\ref{fig:wireCERN}. 

\begin{figure}[h]
\begin{center}
\includegraphics[width=0.8\textwidth]{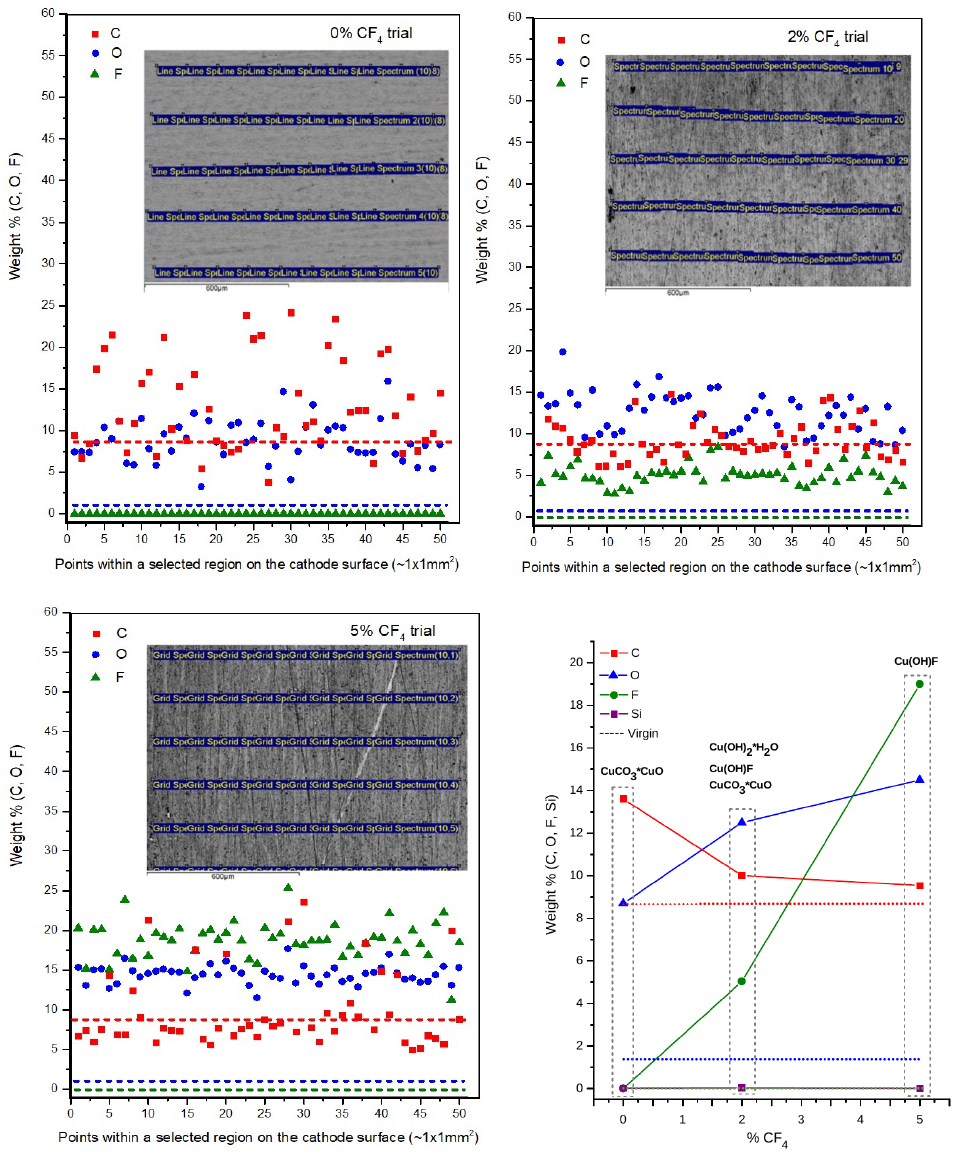}
\vspace{-3 cm}
\caption{Relative weight fraction of carbon, oxygen and fluorine on $1\times1$~mm$^2$ area from the irradiation zone of 0\% (top left), 2\% (top right) and 5\% (bottom left) cathodes measured in point across samples as shown in the SEM images in the top of each plot. The values measured from the virgin cathode sample are provided as dashed lines. The resulting averaged values are shown as function of the  CF$_{4}$ fraction in the gas mixture together with the XRD results (bottom right). 
    \label{fig:EDScathode}}
\end{center}
\end{figure}

The EDS technique is considered to be semi-quantitative but suitable for comparative studies. 
To obtain reliable comparisons of different samples, a multi-point EDS analysis approach was introduced. The final EDS measurements were done in sets of 50 points equally distributed over a sample area of $1200\times1200$~$\mu$m$^2$. This allows measurement of the average content of detected elements on the electrode surface and thus enables comparative studies. This method is used to study the distribution of deposits over the irradiated area and to compare samples from different longevity trials. 

Figure~\ref{fig:EDSwires} shows the results of such analysis applied to anode wires. The left plot shows the averaged relative weight fraction of gold measured for different irradiated wires and compared to the average value obtained for the virgin one. The presence of other elements can be identified as the difference between the gold fractions measured for the virgin and an irradiated wire. The anodes from  2\% and 0\% CF$_{4}$ longevity tests are clearly seen as more strongly affected than the one from the 5\% CF$_{4}$ trial. The carbon distribution along the wires is shown in the middle plot and has a prominent peak in the irradiation area of the 2 and 0\% CF$_{4}$ samples. The results for carbon, silicon, oxygen, fluorine and tungsten were averaged over the wire length and are shown in Fig.~\ref{fig:EDSwires}, right, as functions of the CF$_{4}$ content of the gas mixture.  

A similar approach was applied to the cathode samples. The area of $1\times1$~mm$^2$ from the centre of irradiation zones were measured in 50 points, as shown in Fig.~\ref{fig:EDScathode} (top left, top right, bottom left).  In addition to EDS measurements, X-ray diffraction (XRD) techniques were used to analyze chemical structure of possible crystalline deposits. Figure~\ref{fig:EDScathode} (bottom right) shows the resulting averaged weight fractions for carbon, oxygen, fluorine and silicon as function of the  CF$_{4}$ together with the XRD results. Hydroxide and hydrate groups may indicate presence of water molecules in the gas mixture during the irradiation as a consequence of Rylsan pipes of the gas supply. Carbon concentration is found to be the highest for the 0\% CF$_{4}$ sample, while fluorine presence increases with  CF$_{4}$ fraction. In the XRD analysis carbon was found in CuCO$_3$*CuO with small contributions of graphene oxide phases, and fluorine was identified in Cu(OH)F compound.

%%%%%%%%%%%%%%%%%%%%%%%%%%%%%%%%%%%%%%%%%%%%%%%%%%%%%%%%%%%%%%%%%%%%%%%%%%%%%%%%%%%%%%%%%%%%%%%%%%%%%%%%%%%%%%
% Photos of irradiated cathodes, anodes
%\subsection{Photos of Opened miniCSCs for the 10\% CF$_{4}$ Trial at PNPI}

%%%%%%%%%%%%%%%%%%%%%%%%%%%%%%%%%%%%%%%%%%%%%%%%%%%%%%%%%%%%%%%%%%%%%%%%%%%%%%%%%%%%%%%%%%%%%%%%%%%%%%%%%%%%%%
% SEM, EDS, XEM, etc. analyses of electrodes
%\subsection{Studies of Cathode, Anode Samples by Belgrade Group (10, 5, 2, 0\% CF$_{4}$ Trials)} \label{analysisElectrodesBelgrade}

%Samples were cut from the anodes and cathodes of the irradiated miniCSC and to send to %our colleagues in he Institute of General and Physical Chemistry 
%University of Belgrade.
%There, several techniques were used to examine 
%of the irradiated surfaces, including Scanning Electron Microscopy (SEM), Energy Dispersive Spectroscopy (EDS), X-Ray Diffraction Analysis (XRD), Fourier Transform Infrared Spectroscopy (ATR-FTIR). With these techniques
%we obtain information on the surface morphology, elemental composition of deposits on cathodes and anodes, molecular composition of these deposits (both crystalline and amorphous).

%%%%%%%%%%%%%%%%%%%%%%%%%%%%%%%%%%%%%%%%%%%%%%%%%%%%%%%%%%%%%%%%%%%%%%%%%%%%%%%%%%%%%%%%%%%%%%%%%%%%%%%%%%%%%%

%% file: Conclusion.tex
\section{Summary} \label{conclusion}

Aging proprieties of the CMS Cathode Strip Chambers (CSCs) were studied using miniCSCs, small prototypes built by using the same materials as for the production chambers. 
%small prototypes constructed from the same materials as the production chambers. 
The main purpose was to explore the possibilities of reducing the fraction of CF$_{4}$ in the gas mixture with respect to the nominal 10\% CF$_{4}$.  Longevity studies with 5\%, 2\%, and 0\%  CF$_{4}$ were performed up to the accumulated charge of 300 mC/cm. Additional tests were done with similar prototypes with uniform irradiation and 2\% CF$_{4}$, and with the nominal gas mixture and local irradiation up to 1.2 C/cm. No noticeable detection performance degradation was seen in any of those tests.

However, the electron material analysis done after the longevity tests showed a clear difference in presence of carbon and fluorine on anode and cathode surfaces for irradiation trials with different CF$_{4}$ fractions. An increase in carbon concentration on anode wire surfaces was clearly seen for the samples from  2\%, and 0\%  CF$_{4}$ tests. Silicon presence on anode wires and carbon contamination on cathode surfaces was observed for 0\%  CF$_{4}$. Fluorine was found on cathode samples in quantities increasing with the CF$_{4}$ fraction in the gas. 

Although no longevity issues were observed during the irradiation, the material analysis indicates that the gas mixtures containing less or equal to 2\% of CF$_{4}$ are potentially dangerous for long-term CSC operation due to carbon deposition on anode wires.